\documentclass[10pt,twocolumn,journal]{IEEEtran}
\usepackage{latexsym}
\usepackage{float}
\usepackage{amsfonts}
\usepackage{amsbsy}
\usepackage{amssymb}
\usepackage{times}
\usepackage{graphicx}
\usepackage{setspace}
\usepackage{enumerate}
\usepackage[usenames]{color}
\usepackage[dvips]{pstcol}
\usepackage{epstopdf}
\usepackage[caption=false]{subfig}
\usepackage{cite}
\usepackage{amssymb}
\usepackage{amsfonts}
\usepackage{graphicx}
\usepackage{epsfig}
\usepackage{psfrag}
\usepackage{xcolor}
\usepackage{amsfonts, bm}
\usepackage{epstopdf}
\usepackage{cite}
\usepackage{color}
\usepackage{xcolor}
\usepackage{subfig}
\usepackage{verbatim}
\usepackage{multirow}
\usepackage{booktabs}
\usepackage{amsthm}
\usepackage{makecell}
\usepackage{units}
\usepackage[linesnumbered, ruled]{algorithm2e}
\usepackage{algpseudocode}
\usepackage{amsmath}

\usepackage{array}

\IEEEoverridecommandlockouts
\begin{document}
	\title{A Unified Multi-Task Semantic Communication System for Multimodal Data}
	\author{ Guangyi Zhang, \textit{Graduate Student Member, IEEE}, Qiyu Hu, \textit{Student Member, IEEE}, Zhijin Qin, \textit{Senior Member, IEEE},  Yunlong Cai, \textit{Senior Member, IEEE}, Guanding Yu, \textit{Senior Member, IEEE}, and Xiaoming Tao, \textit{Senior Member, IEEE}
		
		\thanks{Parts of this paper were presented at the IEEE Global Communications Conference, Rio de Janeiro, Brazil, December 2022.
			
			This work was supported in part by the National Natural Science Foundation of China under Grant U22A2004, 62293484, and 61925105, and also in part by Zhejiang Provincial Key Laboratory of Information Processing, Communication and Networking (IPCAN), Hangzhou 310027, China.  \textit{(Corresponding author: Yunlong Cai.)}}
		
		\thanks{ G. Zhang, Q. Hu, Y. Cai, and G. Yu are with the College of Information Science and Electronic Engineering, Zhejiang University, Hangzhou 310027, China (e-mail: zhangguangyi@zju.edu.cn; qiyhu@zju.edu.cn; ylcai@zju.edu.cn; yuguanding@zju.edu.cn).

		Z. Qin is with the Department of Electronic Engineering, Tsinghua University, Beijing 100084, China, and also with the Beijing National Research Center for Information Science and Technology, Beijing 100084, China (e-mail: qinzhijin@tsinghua.edu.cn). 
		
		X. Tao is with the Department of Electronic Engineering, Tsinghua University, Beijing 100084, China (e-mail: taoxm@tsinghua.edu.cn). 
		  } }

	\maketitle
	\vspace{-3.3em}
	\begin{abstract}
		Task-oriented semantic communications have achieved significant performance gains. However, the employed deep neural networks in semantic communications have to be updated when the task is changed or multiple models need to be stored for performing different tasks. To address this issue, we develop a unified deep learning-enabled semantic communication system (U-DeepSC), where a unified end-to-end framework can serve many different tasks with multiple modalities of data. As the number of required features varies from task to task, we propose a vector-wise dynamic scheme that can adjust the number of transmitted symbols for different tasks. Moreover, our dynamic scheme can also adaptively adjust the number of transmitted features under different channel conditions to optimize the transmission efficiency. Particularly, we devise a lightweight feature selection module (FSM) to evaluate the importance of feature vectors, which can hierarchically drop redundant feature vectors  and significantly accelerate the inference. To reduce the transmission overhead, we then design a unified codebook for feature representation to serve multiple tasks, where only the indices of these task-specific features in the codebook are transmitted. According to the simulation results, the proposed U-DeepSC achieves comparable performance to the task-oriented semantic communication system designed for a specific task but with significant reduction in both transmission overhead and model size.
	\end{abstract}
	
	\IEEEpeerreviewmaketitle
	
	\section{Introduction}
	
	With the rapid development of artificial intelligence, a huge amount of interconnected intelligent applications have appeared in the networks. To support massive connectivity for these applications over limited wireless resources, the conventional communication systems are facing critical challenges\cite{6G,6GNetwork,Principle}. To address this issue, semantic communications have been considered as a promising technology to achieve better performance\cite{SemanMaga}.  
	
	\subsection{Prior Work}
	Semantic communications have recently received great attention \cite{MagzineSeman}. Different from conventional communications, they only take into account the relevant semantic information to the tasks \cite{NiuKai, SemEmpower,GoalOriented}. With the integration of communications and deep learning (DL), task-related semantic information can be extracted from the source data through deep neural networks (DNNs), and is represented by the encoded features. Recent DL-based studies on semantic communications have shown a great potential to achieve performance gains \cite{DeepSC, wit, Vedio3, wenzhenzi, Mingyu_TCCN,Constellation,Aini_WCL, Haijun_TCOM, danlan_image,device,Genera, Vedio1,ImagRetri, TransIoT, vqvae,Task-oriented,  JSCCf,Jiaowei_JSAC, Shuai_TWC,Yujie_COML,Jianhao_IOTJ}, especially in unfriendly channel environments.
	
	The existing works on semantic communications can be mainly divided into two categories: data reconstruction \cite{ DeepSC, wit, Vedio3, wenzhenzi, Mingyu_TCCN,Constellation,Aini_WCL, Haijun_TCOM, danlan_image,device,Genera, Vedio1} and task execution \cite{ImagRetri, TransIoT, vqvae,Task-oriented,  JSCCf,Jiaowei_JSAC, Shuai_TWC,Yujie_COML,Jianhao_IOTJ}. For the data reconstruction, the semantic system extracts global semantic information from the source data. Specifically, a so-called DeepSC framework in \cite{DeepSC} encodes the text information into various lengths by employing sentence information. A novel semantic-preserving compression method for text transmission has been developed in \cite{TextCompression}, which saves the number of bits for message representation. In \cite{wit},  an attention-based JSCC system operates with different signal-to-noise (SNR) levels during image transmission. As for video transmission, a DL-aided wireless video transmission system in \cite{Vedio3} can overcome the cliff-effect. The semantic communication system in \cite{wenzhenzi} converts speech signals into semantic features and decodes the received features into a reconstructed speech waveform. For the task execution applications, only the task-specific semantic information is extracted and encoded at the transmitter \cite{ImagRetri, TransIoT, vqvae, Task-oriented,JSCCf}. In particular, a model for image retrieval task under power and bandwidth constraints has been proposed in \cite{ImagRetri}. In \cite{TransIoT}, an image classification-oriented semantic communication system has been developed. In \cite{vqvae}, a vector quantization-variational autoencoder (VQ-VAE) based robust semantic communication systems has been developed for image classification. 
	
	
	Though the aforementioned semantic communication systems have exhibited satisfactory performance in certain scenarios, they only handle one task with single modality of data. These DNN models are hard to simultaneously serve different tasks with multi-modality in practice for the reasons below: (i) The model has to be updated once the task is changed, which leads to massive gradient transmission for retraining it; (ii) Different models need to be stored for serving different tasks, which might be unrealistic for the edge devices with limited storage resources. Generally, most of the devices require multi-task service, hence developing a unified multi-task semantic communication system is of great importance. In \cite{Task-oriented}, a Transformer-based framework has been proposed to address this issue initially. It is able to share the same transmitter structure for the considered tasks. However, the model in \cite{Task-oriented} still needs to be retrained separately for different tasks and the transceiver architecture has not been unified for different tasks yet. A recent work in \cite{MTLS} has designed a model to handle the image detection and segmentation tasks, but it only handles two tasks with one modality of data.  In \cite{MultiReceiver}, a multi-task semantic approach was proposed for joint optimization of completing multiple tasks with multiple receivers, where the considered tasks are mainly image classification tasks.

	There have been some works about multi-task learning in the field of computer vision and natural language processing\cite{Muli-task1,Muli-task2, Unit,mul_r}. Multi-task learning aims at utilizing the task-specific information contained in the training samples of related tasks. Compared with the single-task models, the multi-task models bring the following advantages: (i) The memory space for storing the model can be significantly reduced due to the shared model parameters for multi-task; (ii) It is easier to simultaneously train the model for multiple tasks and improve the performance if some related tasks share the complementary semantic information. Moreover, for data of multiple modalities, multi-representation learning is a significant task, which aims to cope with the consistency and the difference in different modalities of data, and explore modality representations to contain both consistent and complementary information \cite{mul_r}.

	\subsection{Motivation and Contributions}

	It is foreseen that wireless networks are expected to provide various intelligent services in terms of generality and efficiency. Though there have been numerous semantic works for specific tasks, a unified multi-task model for different modalities of data in wireless communications has not been thoroughly investigated. Therefore, in this paper, we propose a unified DL-based semantic communication system (U-DeepSC). To the best of our knowledge, this is the first work on a unified semantic communication system designed to serve various tasks.
	Furthermore, it is important to notice that there is redundancy in transmitting all the features, as different tasks require varying numbers of transmitted features. For instance, data reconstruction typically demands more transmitted features compared to intelligent tasks. While transmitting more encoded features can enhance performance against noise by capitalizing on feature redundancy, this also introduces higher transmission overhead. Consequently, an inherent trade-off between performance and the number of transmitted symbols exists.
	Nevertheless, the majority of existing methods heavily rely on handcrafted designs, where the output size of employed DNN models remain fixed. This will cause the transmission rate to exceed its required minimum value. Therefore, it is important to determine an optimal transmission rate for each task within a unified multi-task semantic communication system.

	To this end, we make the first attempt to devise a unified semantic communication for end-to-end data transmission. Our approach takes into account six widely recognized tasks within the semantic communication community, encompassing both single-modality tasks and multi-modality tasks. The proposed U-DeepSC is capable of simultaneously handling multiple tasks across three distinct modalities: image, text, and speech. In order to equip U-DeepSC with the ability to extract task-specific information for diverse tasks, we introduce task embedding vectors and task query matrices. These components are input alongside the source data for each task, effectively indicating the intended task for the given data. To determine an appropriate transmission overhead tailored to each task, we develop a dynamic channel encoder for U-DeepSC. This encoder is designed to drop redundant feature vectors  for specific tasks and adjust the number of transmitted feature vectors based on the channel conditions. More specifically, we implement a lightweight feature selection module (FSM) within the channel encoder. This module generates a selection mask vector, taking into account both the tasks and current channel conditions, to precisely indicate which features need to be transmitted. In addition, the proposed FSM can evaluate the importance of feature vectors and hierarchically prune redundant feature vectors, significantly speeding up the inference.  To reduce the transmission overhead as well as enable digital transmission, we adopt the codebook design in \cite{vqvae}, where a discrete codebook shared by the transmitter and receiver is designed for encoded feature representation and only the indices of these encoded features in the codebook are transmitted. Different from \cite{vqvae} where the codebook is for a specific task with single modality, we design a unified codebook for multimodal data. Furthermore, we develop a unified decoder for different tasks, where a masked cross-attention method is proposed for parallel training. Simulation results show that our proposed methods achieve comparable performance to the task-oriented semantic communication systems designed for a specific task with much reduced transmission overhead and fewer model parameters. 
	
	Specifically, the main contributions of this paper can be summarized as follows.

	\begin{itemize}
		\item We propose a unified semantic communication framework, U-DeepSC, which can handle a number of tasks using a fixed model.
		The proposed U-DeepSC is a general framework that can support the transmission of three modalities of data.
		
		\item We design FSM at the channel encoder, which is empowered with the ability to dynamically adjust the numbers of transmitted features under different channel conditions and tasks. It is able to achieve an adaptive tradeoff between transmission rate and task performance for the considered tasks. Since FSM can hierarchically prune redundant feature vectors, the computation complexity is reduced and inference speed can be accelerated. 
		
		\item We develop a unified codebook for multi-task services to support digital communication and reduce transmission overhead. Specifically, we introduce vector quantized-variational mechanism for discrete feature representations, along with the utilization of a digital modulation module for digital transmission.
		
		\item We devise a unified decoder based on Transformer decoder, where a novel masked cross-attention method is proposed to achieve parallel training. Besides, we also propose a novel two-phase training algorithm to simultaneously learn multiple tasks.

	\end{itemize}

	\subsection{Organization and Notations}
	The rest of this paper is organized as follows. Section II introduces the framework of U-DeepSC and the considered tasks. Section III presents the detailed architecture of semantic encoder. Section IV introduces the dynamic channel encoder and FSM. The unified codebook and decoder are introduced in Section V. The masked cross-attention for parallel training and training method are provided in Section VI. Simulation results are presented in Section VII. Finally, Section VIII concludes this paper.
	
	\emph{Notations:} Scalars, vectors, and matrices are respectively denoted by lower case, boldface lower case, and boldface upper case letters. For a matrix $\boldsymbol{A}$, $\boldsymbol{A}^\top$ and $\|\boldsymbol{A}\|$ denote its transpose, conjugate, and Frobenius norm, respectively. For a vector $\boldsymbol{a}$, $\|\boldsymbol{a}\|$ is its Euclidean norm. 
	Additionally, $\odot$ is the element-wise multiplication of two matrices, i.e., Hadmard product. 
	Finally, ${\mathbb{C}^{m \times n}}\;({\mathbb{R}^{m \times n}})$ are the space of ${m \times n}$ complex (real) matrices. Most of the key notations are listed in Table I.

	\begin{table}[t]\footnotesize
		\centering
		\label{NotationList}
		\caption{\fontsize{10pt}{17bp}Definitions of Notations.}  
		\begin{tabular}{l | l }
			\toprule[0.3mm]
			Notation & Definition  \\
			\midrule[0.18mm]
			\midrule[0.18mm]
			$\bm{s}^v$, $\bm{s}^t$, $\bm{s}^s$ & input image, text,  and speech signals\\
			$L_v$, $L_t$, $L_s$ & length values of image, text, and speech signals \\
			$\bm{z}^v$, $\bm{z}^t$, $\bm{z}^s$ & encoded input image, text,  and speech features\\
			$K_v$, $K_t$, $K_s$ & numbers of channel uses  \\
			$\bm{y}^v$, $\bm{y}^t$, $\bm{y}^s$ & received input image, text,  and speech features\\
			$\bm{\theta}_v$, $\bm{\theta}_t$, $\bm{\theta}_s$ & trainable  parameters of transmitters \\
			$f_v$, $f_t$, $f_s$ & mappings  of  image, text,  and speech transmitter\\
			$\bm{\phi}_u$, $f_u$ & trainable  parameters and mapping of receiver\\
			$h^i,\bm{n}^i$, $i\in \{v,t,s\}$ & channel gain coefficients and noises \\
			$\bm{X}^v,\bm{X}^t,\bm{X}^s$ & preprocessed feature matrices \\
			$E_v$, $E_t$, $E_s$ & length values of feature vectors in  $\bm{X}^v,\bm{X}^t,\bm{X}^s$\\
			$N_v$, $N_t$, $N_s$ & numbers of feature vectors in  $\bm{X}^v,\bm{X}^t,\bm{X}^s$\\
			$\bm{w}^v_{tk},\bm{w}^t_{tk},\bm{w}^s_{tk}$ & task-specific embedding vectors\\
			$\hat{\bm{X}}^v,\hat{\bm{X}}^t,\hat{\bm{X}}^s$ & feature matrices with $\bm{w}^v_{tk},\bm{w}^t_{tk},\bm{w}^s_{tk}$\\
			$\bm{X}^v_{se},\bm{X}^t_{se},\bm{X}^s_{se}$  & output feature matrices of semantic encoder \\
			$\bm{X}^v_{en},\bm{X}^t_{en},\bm{X}^s_{en}$ & output feature matrices of channel encoder \\
			$\bm{F}_g,\bm{F}_l,\bm{F}_n$ & global, local, noise feature mapping matrices\\
			$\bm{F}_c$ & comprehensive feature matrix\\
			$\bm{F}^q,\bm{m}^q$ &$q$-th  feature matrix and $q$-th mask vector \\
			$N_d$ &number of feature vector in $\bm{F}^q$\\
			$\hat{\bm{m}}$ & final mask vector\\
			$\bm{P}$, $\bm{o}$ & sampling probability matrix and one-hot vector\\
			$\bm{e}_j^i, i \in \{v,t,s\}$ & $j$-th basis vector of codebooks\\
			$N_q$ & number of FSM\\
			$f_{vq}$ & mapping for selecting basis vectors in codebook\\
			$f_{n}$ & mapping for prefined masking ratio\\
			$\delta$ & predefined ratio if transmitted feature vectors   \\
			$\bm{U}_{cd}$ & feature matrix after channel decoder\\
			$\bm{U}_{de}^{i}$ & feature matrix at $i$-th semantic decoder layer\\
			$\bm{W}_{ry}$ & task-specific query matrix\\
			$\bm{A},\bm{M}$ & attention score and attention mask  matrices\\
			$\bm{B}$  &  cross-attention matrix\\
			\bottomrule[0.3mm]
		\end{tabular}
	\end{table}

	\begin{figure*}[t]
		\begin{centering}
			\includegraphics[width=0.97 \textwidth]{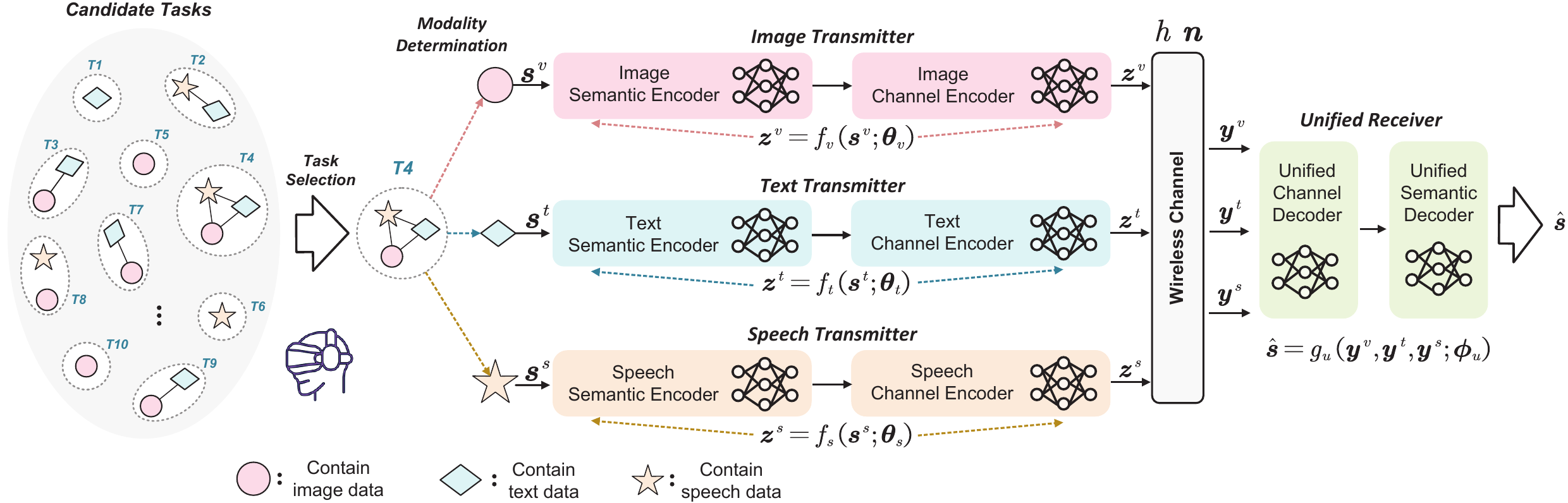}
			\par \end{centering}
		\caption{Framework of the proposed U-DeepSC.}
		\label{FrameArchitect}
	\end{figure*}
	
	\section{Framework of Unified Semantic Communication} \label{framework}
	In this section, we design the framework of U-DeepSC. The U-DeepSC consists of the semantic/channel encoders for each modality, and the unified semantic/channel decoder with light-weight task-specific heads.
	
	\subsection{System Model}
	As shown in Fig. \ref{FrameArchitect}, the proposed U-DeepSC is able to handle a number of tasks with three modalities. We assume that there is a candidate task set that contains various tasks, e.g. $\textrm{T}1, \textrm{T}2,..., \textrm{T}10$. These tasks may involve different numbers of modalities, including image, text, and speech. We then take T$4$ that contains image, text, and speech data as an example to present the overall process of U-DeepSC. In particular, after determining the involved modality, the proposed framework will activate the corresponding transmitters based on the selected modalities. It mainly consists of four parts: an image transmitter, a text transmitter, a speech transmitter, and a unified receiver. DNNs are employed to implement the transmitters and the unified receiver. In particular, each transmitter consists of a semantic encoder and a dynamic channel encoder. The unified receiver consists of the channel decoder and the semantic decoder. 
	
	The inputs of the system are image $\boldsymbol{s}^{v}\! \in \!\mathbb{R}^{L_v}$, text $\boldsymbol{s}^{t} \! \in \! \mathbb{R}^{L_t}$, and speech $\boldsymbol{s}^{s} \! \in \! \mathbb{R}^{L_s}$, where $L_v$, $L_t$, and $L_s$ are the length values of image, text, and speech signals, respectively. Specifically, $L_v$ is the size of an image, $L_t$ signifies  the word count in a sentence, and $L_s$ is the number of sampling points in the speech signal. Subsequently, each of these signals undergoes processing to obtain the corresponding encoded features. For simplicity in notation, we treat the semantic encoder and channel encoder collectively.  In particular, each signal undergoes initial encoding by the semantic encoder to extract semantic features.  Subsequently, the dynamic channel encoder is designed to execute feature selection, regulating the size of the channel symbols.  As a result, the encoded features post the channel encoders for the three data modalities are denoted as $\boldsymbol{z}^{v} \! \in \! \mathbb{C}^{K_v}$, $\boldsymbol{z}^{t} \! \in \! \mathbb{C}^{K_t}$, and $\boldsymbol{z}^{s} \! \in \! \mathbb{C}^{K_s}$, where $K_v$, $K_t$, and $K_s$ denote the numbers of channel uses of image, text, and speech signals, respectively.  The encoding process can be expressed as 
	\begin{equation}
		\bm{z}^i = f_i(\bm{s}^i;\bm{\theta}_{i}), i \in \{v,t,s\},
	\end{equation}
	where $f_v(\cdot)$, $f_t(\cdot)$, and $f_s(\cdot)$ denote the mappings of image, text, and speech transmitters, respectively. Additionally, $\bm{\theta}_{v}$, $\bm{\theta}_{s}$, and $\bm{\theta}_{t}$ are the trainable parameters in image, text, and speech transmitters, respectively.
	We define the bandwidth ratios for three modalities as $\rho_{i}\! \triangleq \! K_{i}/L_{i} , i \! \in \! \{v,t,s\}$. Moreover, the transmitted signals are subject to a power constraint $P$, i.e., 
	$\frac{1}{K_i}\mathbb{E}\|\boldsymbol{z}^i\|^2 \leq P$, $i \in \{v,t,s\}$.

	The encoded vectors, $\bm{z}^i $, $i \in \{v,t,s\}$,  are then sent to the receiver for decoding, and the received vectors are denoted as $\boldsymbol{y}^i, i \in \{v,t,s\}$.
	Specifically, the received vector at the receiver is given by 
	\begin{equation}
		\boldsymbol{y}^i = h^i \boldsymbol{z}^i + \boldsymbol{n}^i, i \in \{ v,t,s \},
	\end{equation}
	where $h^i \in \mathbb{C}$ represents the channel gain  coefficient  and $\boldsymbol{n}^i\sim \mathcal{CN}(\bm{0}, \sigma^{2})$ is the additive white Gaussian noise (AWGN). 
	
	At the receiver, the decoder firstly processes the corrupted complex-valued signal and the obtained features are further processed by light-weight task-specific head to execute downstream tasks. The decoding process can be formulated as 
	\begin{equation}
		\hat{\bm{s}} = f_u(\bm{{y}}^v,\bm{{y}}^t,\bm{{y}}^s; \bm{\phi}_u),
	\end{equation}
	where $f_u(\cdot)$ denotes the mapping of the receiver, including channel and semantic decoder, and $\bm{\phi}_u$ is the trainable parameters of the receiver. Moreover, as different tasks may involve different modalities of data, there might not always be a necessity to simultaneously feed all three modalities of data into $f_u(\cdot)$. For instance, for tasks like image classification that solely require image data, only $\bm{y}^v$ is necessary.

	\subsection{Task Description} 
	
	To conduct a comprehensive analysis of U-DeepSC and present compelling results illustrating its efficacy, we will address a diverse set of significant tasks across various domains. These tasks encompass text classification, visual question answering, image classification, video sentiment analysis, image data reconstruction, and text data reconstruction. They involve three fundamental data modalities: image, text, and speech, which constitute a substantial component of contemporary multimedia communication. Additionally, we include cross-modality tasks such as visual question answering, aligning with prevalent themes in existing semantic communication research. The selected tasks serve as a representative demonstration of the proposed scheme's effectiveness, with their relevance evident in established semantic communication systems \cite{DeepSC,Task-oriented,ImagRetri,TextCompression}.  
	\begin{itemize}
		
		\item[(i)]
		\textbf{Text classification:}
		The purpose of the text classification task is to classify whether the sentiment of a given sentence is positive or negative. It is essentially a binary classification problem. Thus, we take classification accuracy as the performance metric for text classification, and the cross-entropy as the loss function to train the model. 
		
		\item[(ii)]
		\textbf{Visual question answering:}
		In visual question answering task, the images and questions in text are processed by the model to classify which answer is right. Thus, we take answer accuracy as the performance metric and the cross entropy as the loss function. 
		
		\item[(iii)]
		\textbf{Video sentiment analysis:}
		The video sentiment analysis task is about leveraging multimodal signals for an effective understanding of the videos generated by users \cite{video}. As for the evaluation criterion, classification accuracy is selected as the metric. Additionally, the binary cross-entropy is used for the loss function.  
		
		\item[(iv)]
		\textbf{Image classification:}
		The image classification task aims at classifing which category the given image belongs to. To evaluate the performance of image classification task, the classification accuracy is adopted as the performance evaluation metric. To learn this task, we adopt the cross-entropy loss function.
		
		\item[(v)]
		\textbf{Image reconstruction:}
		The performance of the image reconstruction task is quantified by the peak signal-to-noise ratio (PSNR). The PSNR measures the ratio between the maximum possible power and the noise, which is given by
		\begin{equation}
			\textrm{PSNR}=10 \log_{10}{\frac{\textrm{MAX}^{2}}{\textrm{MSE}}}(\textrm{dB}),
		\end{equation}
		where $\textrm{MSE}=d(\boldsymbol{x},\hat{\boldsymbol{x}})$ denotes the mean square error (MSE) between the source image $\boldsymbol{x}$, and the reconstructed image $\hat{\boldsymbol{x}}$, and $\textrm{MAX}$ is the maximum possible value of the pixels. Moreover, the MSE is adopted as the training loss.

		\item[(vi)]
		\textbf{Text reconstruction:}
		As for the text reconstruction task, the bi-lingual evaluation understudy (BLEU) score is adopted to measure the performance.  The BLEU takes the n-gram matching criterion to measure the performance. 
		BLEU score is a scalar between $0$ and $1$, which evaluates the similarity between the reconstructed text and the source text, with $1$ representing the highest similarity. We take the cross entropy as the loss function since the BLEU score is non-differentiable.
		
	\end{itemize}

	\section{Semantic Encoding Design for Multimodal Data}
	In this section, we elaborate the detailed design of the transmitter, including the semantic encoders for three modalities of data. Since the data from different modalities have totally different statistical characteristics and have different semantic information, we have to design image, text, and speech semantic encoders for image, text, and speech, respectively. The detailed architecture of the transmitter is shown in Fig. \ref{Transmitter}.
	Specifically, the overall process of transmitter is depicted as
	\begin{equation}
		\begin{split}
			&\boldsymbol{s}^i {\xrightarrow[]{}} \boldsymbol{X}^i {\xrightarrow[]{\boldsymbol{w}_{tk}^i}} \hat{\boldsymbol{X}}^i \xrightarrow{}  \boldsymbol{X}^i_{se} \xrightarrow[\textit{}]{\boldsymbol{m}^q}  \boldsymbol{X}^i_{en},\;    i \in \{ v,t,s \},
		\end{split}
	\end{equation}
	where the definitions of notations can be found in the following sections.

	\subsection{Transformer-Based Encoding}
\begin{figure}[t]
		\begin{centering}
			\includegraphics[width=0.48\textwidth]{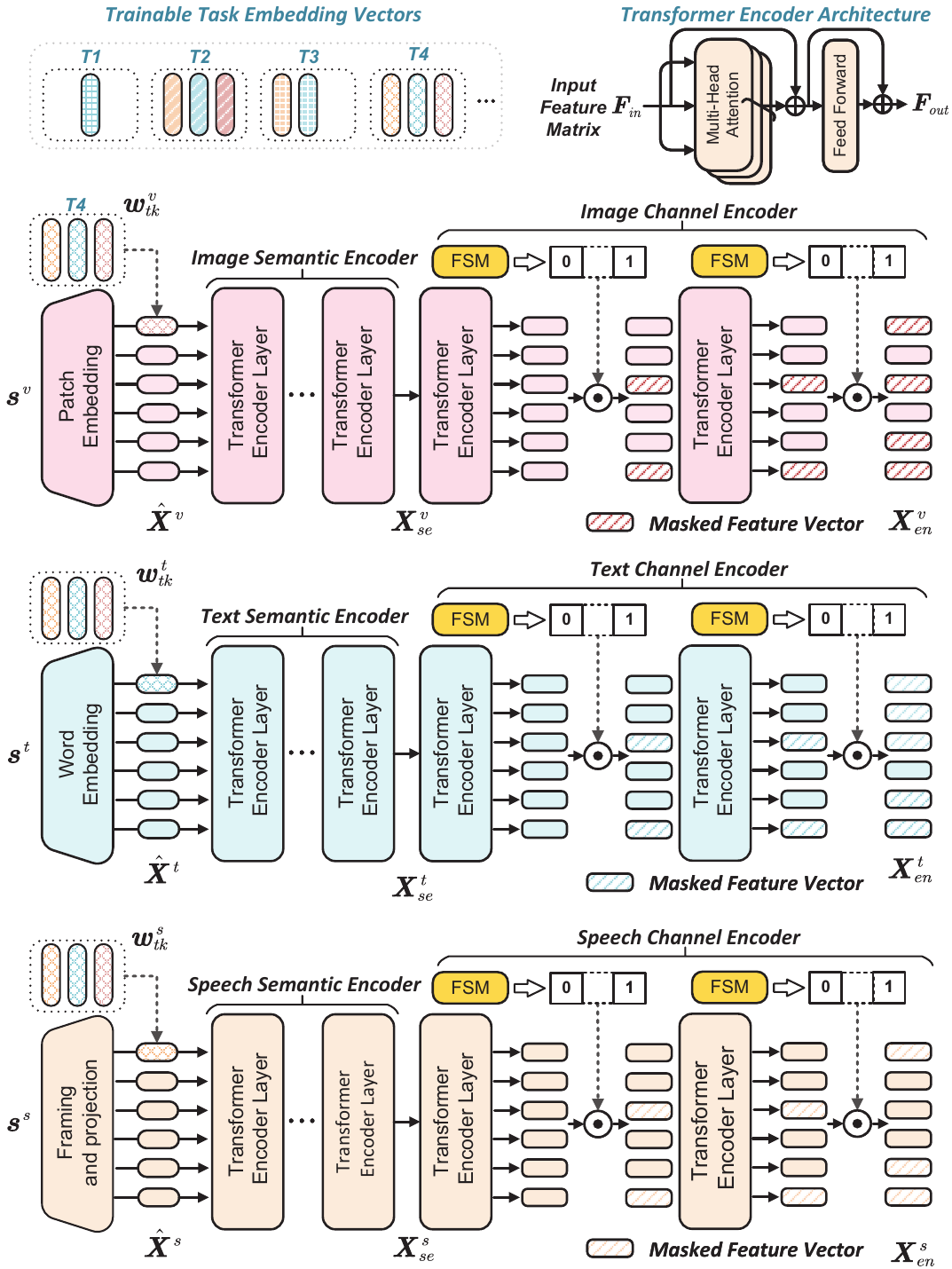}
			\par\end{centering}
		\caption{The transmitter architecture of the proposed U-DeepSC. It mainly consists of three transmitters and each transmitter is comprised of two parts: semantic encoder and channel encoder. The task embedding vectors consist of randomly generated trainable parameters, which are continually updated along with the model. Additionally, multiple tasks will share a common transmitter.} 
		\label{Transmitter}
	\end{figure}

	We mainly use Transformer encoder to construct the encoders of the transmitter.
	The Transformer encoder constitutes the fundamental building block in the landscape of signal processing \cite{Bert,ViT,backbone}. 
	The architecture of a basic Transformer encoder layer is depicted in the upper part of Fig. \ref{Transmitter}. Each Transformer encoder layer, as illustrated in Fig. \ref{Transmitter}, comprises a multi-head self-attention (MSA) module and a multi-layer perceptron (MLP) layer. The MSA mechanism enables the model to capture long-range dependencies among different feature vectors, facilitating efficient global information integration. It assigns higher weights to relevant feature vectors while simultaneously attending to different positions in the sequence. After the MSA, the feed forward neural (FFN) layer introduces non-linearity and captures complex relationships within the feature vectors.
	
	As  depicted  in Fig. \ref{Transmitter}, we  input  the intermediate
	feature matrix $\bm{F}_{in}\in  \mathbb{R}^{L\times D}$ into the Transformer encoder layer. The generated feature matrix is represented by $\bm{F}_{out} \in \mathbb{R}^{L\times D} $, and both $\bm{F}_{in}$ and $\bm{F}_{out}$ share the shape $L\times D$.
	Overall, the encoding procedure of a Transformer encoder layer \cite{ViT} can be denoted as 
	\begin{equation}
		\bm{F}_{out}= \textit{MSA}(\bm{F}_{in})+\textit{FFN}(\textit{MSA}(\bm{F}_{in})).
	\end{equation}
	 Furthermore, GeLU activation and layer normalization operations are employed prior to  MSA and MLP modules. 
	
	The MSA module consists of $N_a$ self-attention modules with a residual connection, i.e.,
	\begin{equation}
		\begin{split}
			& \textit{MSA}(\bm{X}_{in})= \\
			&\bm{F}_{in}+\textit{concat}\underbrace{\left(\textit{SA}(\bm{F}_{in}),\textit{SA}(\bm{F}_{in}),...,\textit{SA}(\bm{F}_{in})\right)}_{N_a} \bm{W},
		\end{split}
	\end{equation}
	where $\textit{concat}(\cdot)$ denotes the concatenation operation,  $\textit{SA}(\bm{F}_{in}) \in \mathbb{R}^{L\times D_a}$ is self-attention operation on $\bm{F}_{in}$,  $\bm{W} \in \mathbb{R}^{D_aN_a\times D}$ is linear projection matrix, and $D_a=D/N_a$ is the output dimension. Moreover, $\textit{SA}(\cdot)$ is formulated as 
	\begin{equation}
		\textit{SA}(\bm{F}_{in}) = \textit{softmax}\left( \frac{ \bm{Q} \bm{K}^{\top}}{\sqrt{D}} \right)\bm{V},
	\end{equation}
	where $\bm{Q} \in \mathbb{R}^{L \times D_a}$, $\bm{K} \in \mathbb{R}^{L \times D_a}$, and $\bm{V} \in \mathbb{R}^{L \times D_a}$ are generated from three linear projections as 
	\begin{equation}
		\bm{Q}=\bm{F}_{in}\bm{W}_Q,\quad\bm{K}=\bm{F}_{in}\bm{W}_K, \quad\bm{V}=\bm{F}_{in}\bm{W}_V,
	\end{equation}
	where $\bm{W}_Q\in \mathbb{R}^{D \times D_a}$, $\bm{W}_K\in \mathbb{R}^{D \times D_a}$, $\bm{W}_V \in \mathbb{R}^{D \times D_a}$, and $\textit{softmax}(\cdot)$ denotes the Softmax function. Moreover, all the introduced projection matrices are set as trainable parameters.
	In addition to MSA, the FFN module consists of a multi-layer MLP and a GeLU activation function.

	\subsection{Semantic Encoder Design}
	\subsubsection{Image Semantic Encoder}
	The image-only and multimodal tasks take an image $\boldsymbol{s}^v$ as input. Then, it is preprocessed \cite{ViT} through patch embedding into the preliminary features $\boldsymbol{X}^{v}=[ \boldsymbol{x}^v_{1},  \boldsymbol{x}^v_{2},\dots, \boldsymbol{x}^v_{N_v}]^{\top} \in \mathbb{R}^{N_v \times E_v}$, where $N_v$ and $E_v$ denote the number and length of the feature vectors, respectively. Additionally, for the video tasks that involve processing multiple frames of images at one time, we concatenate the preliminary features of different frames as the input of the image semantic encoder. U-DeepSC is designed as multi-task model that can simultaneously deal with several tasks with just a set of parameters. Besides, there is no need to update the parameters when the task changes. In this case, the encoder of U-DeepSC needs to identify the current task so that it can perform the feature extraction pertinently.

	To this end, we introduce the task embedding vector to the semantic encoder, which is generated as trainable parameter and updated with the whole model. Specifically, we would generate one task embedding vector for each task at the image encoder. For instance, in the case of two tasks related to image data, two task embedding vectors are randomly generated. It is worth noting that the task embedding vector is quite similar to the CLS token embedding vector in the vanilla transformer \cite{ViT,Bert}. This is achieved by generating a set of random trainable parameter vectors, as illustrated in Fig. \ref{Transmitter}, and jointly learning them across the entire network. During the training phase, we select the task embedding vector corresponding to the specific task and input it along with the image. This action signifies the task to be performed, enabling the semantic encoder to extract task-specific information. In the inference phase, the learned task embedding vectors are selected for specific tasks. For instance, if a task involves three modalities, three learned task embedding vectors, such as T4, are selected. As shown in Fig. \ref{Transmitter}, the task embedding vector $\boldsymbol{w}_{tk}^{v} \in \mathbb{R}^{E_v}$ is added to the semantic encoder. It is  concatenated with $\boldsymbol{X}^v$, and the concatenated feature matrix $\hat{\boldsymbol{X}} \in \mathbb{R}^{(N_v+1)  \times E_v}$ is given  by
	\begin{equation}
		\vspace{0.5ex}
		\hat{\boldsymbol{X}}^v=\textit{concat}\left(\boldsymbol{X}^{v}, \boldsymbol{w}_{tk}^{v}\right).
		\vspace{0.5ex}	
	\end{equation}
	Particularly, the task embedding vectors are employed to perform convolution or attention operation together with the encoded image feature vectors \cite{MAE}. Then, we obtain the encoded image feature matrix through the image semantic encoder, which is represented by $\boldsymbol{X}^{v}_{se}$.

	\subsubsection{Text Semantic Encoder}
	As for text-only and multimodal tasks, we preprocess the input text by word embedding into a sequence of $N_t$ feature vectors, $\boldsymbol{X}^{t}=[ \boldsymbol{x}_{1}^{t}, \boldsymbol{x}_{2}^{t},\dots,\boldsymbol{x}_{N_t}^{t} ]^{\top} \!\in \! \mathbb{R}^{ N_t \times E_t }$, where $N_t$ and $E_t$ denote the number and length of the feature vectors, respectively. Subsequently, $\boldsymbol{X}^{t}$ is encoded by the text semantic encoder. Similar to the image semantic encoder, we also add a trainable task embedding vector $\boldsymbol{w}_{tk}^{t} \in \mathbb{R}^{E_t}$ by concatenating it with $\boldsymbol{X}^{t}$. Then, the concatenated sequence $\hat{\boldsymbol{X}}^t\in \mathbb{R}^{ (N_t+1) \times E_t}$ is input to the text semantic encoder, which generates the encoded text feature matrix $\boldsymbol{X}^t_{se}\in \mathbb{R}^{ (N_t+1) \times E_t}$.

	\subsubsection{Speech Semantic Encoder}
	As for speech-only and multimodal tasks, the input speech signal of the proposed system is obtained from the speech set. Similarly, we first preprocess speech signal $\boldsymbol{s}^s$ into $\boldsymbol{X}^{s} = [ \boldsymbol{x}_{1}^{s}, \boldsymbol{x}_{2}^{s},\dots,\boldsymbol{x}_{N_s}^{s} ]^{\top} \in \mathbb{R}^{ N_s \times E_s }$, where $N_s$ and $E_s$ denote the number and length of the feature vectors, respectively. Then, the speech semantic encoder learns the $N_s$ encoded speech feature matrix $\boldsymbol{X}^{s}_{se}  \in \mathbb{R}^{(N_s+1) \times E_s}$ from the concatenated sequence $\hat{\boldsymbol{X}^{s}}=\textit{concat}\left(\boldsymbol{X}^{s}, \boldsymbol{w}_{tk}^{s}\right) \in \mathbb{R}^{(N_s+1) \times E_s}$, where $\boldsymbol{w}_{tk}^{s} \in \mathbb{R}^{E_s}$ is the task embedding vector trained with the whole network.

	\section{Feature Selection-Based Dynamic Channel Encoder}
	In this section, we introduce the task-specific dynamic overhead by developing the dynamic channel encoder with FSM. 
	
	\subsection{Hierarchical Feature Selection}
	Transmitting all the feature vectors introduces excessive redundancy of semantic information and different tasks require different numbers of transmitted features. Thus, the transmission overhead in U-DeepSC can be reduced by selecting a specific number of feature vectors for each task. Although excessive redundancy generally induces a high transmission overhead and latency, it leads to better performance against noise if more encoded features are transmitted. It is mainly because when certain features are seriously disturbed, the other features that are not disturbed can help to maintain the performance. Therefore, we need to balance the performance and the number of transmitted symbols. To achieve this goal, we design a channel encoder to dynamically adjust the number of output feature vectors for different tasks under different channel conditions in U-DeepSC, which is able to dynamically achieve satisfactory performance by transmitting the least number of features.
	
	Particularly, we design a dynamic channel encoder to conduct vector-wise feature selection  by adjusting the number of the transmitted feature vectors. The proposed dynamic channel encoder has the following advantages:
	\begin{itemize}
		\item The transmitter can identify the task-related features and omit the task-unrelated features, which leads to satisfactory performance.
		
		\item  In Transformer architecture, each encoded feature corresponds to portion of the input, such as one small patch of the input image or one word of a sentence. Thus, it is also of great interpretability to see which features are task-unrelated.
		
		\item With the vector-wise selection design, the transmission overhead can be dynamically adjusted and significantly reduced.

	\end{itemize}

\begin{figure}[t]
		\begin{centering}
			\includegraphics[width=0.45\textwidth]{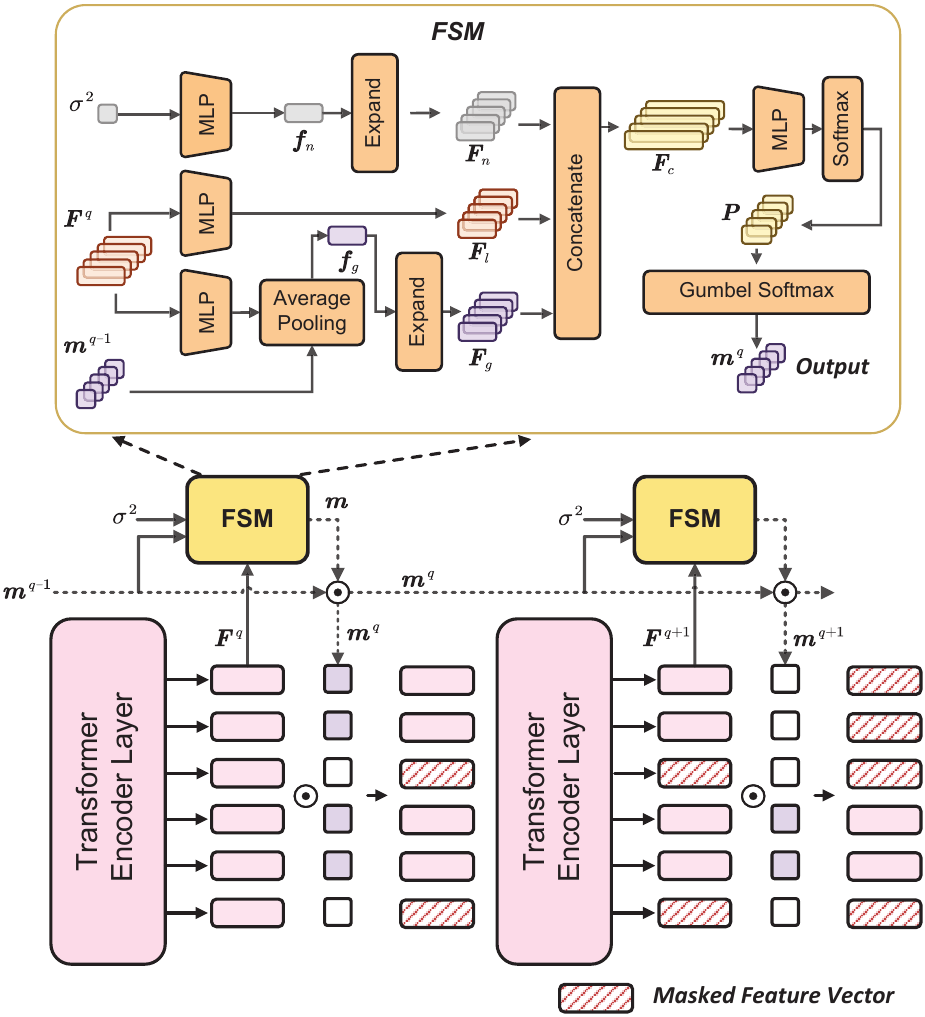}
			\par\end{centering}
		\caption{The architecture of the proposed channel encoder and FSM. The FSM takes $\sigma^2$, $\bm{F}^{q}$, and $\bm{m}^{q-1}$ as input and outputs $\bm{m}^q$.}
		\label{DynaOver}
	\end{figure}

	Since the output of the semantic encoder is a sequence of encoded feature vectors. We omit the task-unrelated features and transmit the informative task-related features to the receiver. In particular, as shown in Fig. \ref{DynaOver}, the FSM is inserted into the original channel encoder layer to generate the mask vector, which is used to indicate the features to be kept/dropped for each task. Inspired by \cite{DynaViT}, we conduct hierarchical selection by inserting multiple FSMs at certain layers, to perform gradually dropping. Moreover, hierarchical selection gradually drops the relatively unimportant features and can avoid mistakenly dropping important features directly. It is equivalent to dividing a complex selection problem into several simple selection problems. After dropping a small part of features each time, the model can adaptively adjust the next selection according to the existing unmasked features, to achieve a higher tolerance for error selection than the single selection performed at the end of the transmitter. 
	
	\subsection{FSM Design}
	\subsubsection{Architecture Details}
	To realize hierarchical feature selection, i.e., gradually drop the uninformative feature vectors, we introduce the selection mask vector, $\boldsymbol{m}^q =[m^q_1,m^q_2,...,m^q_{N_d}]\! \in \! \{ 0,1 \} ^{N_d}$, which indicates whether to transmit each feature vector and $N_d$ is the number of the encoded feature vectors. The elements in the selection mask vector are firstly initialized to $1$ and updated progressively in the $N_q$ FSMs, as shown in Fig. \ref{DynaOver}. We denote the feature matrix $\boldsymbol{F}^{q} =[  \boldsymbol{f}_1^q,\boldsymbol{f}_2^q,\dots,\boldsymbol{f}_{N_d}^q  \big]^{\top} \in \mathbb{R}^{N_d\times E_d}$, as the input of the $q$-th selection module, where $N_d$ and $E_d$ denote the number and the size of the feature vectors, respectively. The $q$-th selection module takes the previous selection mask vector $\boldsymbol{m}^{q-1}$ and $\boldsymbol{F}^{q}$, as input, and outputs the probabilities of keeping the feature vectors. 
	
	Firstly, the encoded feature vectors are projected by a MLP:
	$\boldsymbol{F}_{l}\!=\!\textit{MLP}(\boldsymbol{F}^{q})\in \mathbb{R}^{N_d\times E_l}$, where $\textit{MLP}(\cdot)$ denotes the MLP layer, $\boldsymbol{F}_{l}$ denotes the information feature matrix with a smaller size than the original feature matrix, and $E_l$ is smaller than $E_d$. In addition, to determine whether to drop or keep a feature vector, we also need to consider the information of the existing feature vectors that are not masked. It is employed to perform information fusion with $\boldsymbol{F}_{l}$ to evaluate the information loss caused by the mask operation. Therefore, the global feature vector $\boldsymbol{f}_{g}$ that aggregates the information of all existing feature vectors that have not been masked can be computed by
	\begin{equation}
		\boldsymbol{f}_{g}={g} \big( \textit{MLP} (\boldsymbol{F}^{q}), \boldsymbol{m}^{q-1} \big) \in \mathbb{R}^{ E_l},
	\end{equation}
	where $g(\cdot)$ is the function to extract the information of the existing feature vectors. $g(\cdot)$ can be implemented by the average pooling
	\begin{equation}
		g(\boldsymbol{F}, \boldsymbol{m})={\sum_{i=1}^{N_d} m_i \cdot \boldsymbol{f}_{i}}/{\sum_{i=1}^{N_d} m_i},
	\end{equation}
	where $i$ is the index of the vector, $\boldsymbol{F}=[\boldsymbol{f}_1,\boldsymbol{f}_2,\dots,\boldsymbol{f}_{N_d}]^{\top}$, and $\boldsymbol{m}=[m_1,m_2,\dots,m_{N_d}]^{\top}$. 
	We further  expand $\boldsymbol{f}_{g}$ as $\boldsymbol{F}_{g}=[\boldsymbol{f}_{g}, \boldsymbol{f}_{g},\dots,\boldsymbol{f}_{g}]^{\top}\in \mathbb{R}^{N_d \times E_l}$ to perform knowledge fusion between $\boldsymbol{F}_{g}$ and $\boldsymbol{F}_{l}$. The channel noise  also affects probability of keeping the feature vectors, we employ another MLP layer to extract the noise feature  $\boldsymbol{f}_n=\textit{MLP}(\sigma^2)\in \mathbb{R}^{ E_l}$, where $\sigma^2$ is the noise variance. Then, we expand $\boldsymbol{f}_n$ in the same way as $\boldsymbol{f}_{g}$ to obtain $\boldsymbol{F}_n \!\in \! \mathbb{R}^{ N_d\times E_l}$. As a result, we combine the local, global, and noise feature to obtain comprehensive feature matrix
	\begin{equation}
		\boldsymbol{F}_c=\textit{concat}\left(\boldsymbol{F}_{l }, \boldsymbol{F}_{g}, \boldsymbol{F}_{n}\right) \in \mathbb{R}^{ N_d\times 3E_l},
	\end{equation} 
	and feed it to another MLP to predict the probabilities of keeping the feature vectors,
	\begin{equation}
		\begin{aligned}
			\boldsymbol{P} &=\textit{softmax}(\textit{MLP}(\boldsymbol{F}_c)) \in \mathbb{R}^{N_d \times 2}. 
		\end{aligned}
	\end{equation}
	Then, we can obtain the current mask vector $\boldsymbol{m}$ by sampling from $\boldsymbol{P}$. Note that this module will be inserted into different layers of the channel encoder to perform gradually dropping feature vectors. Accordingly, the true mask vector ${\boldsymbol{m}}^q$ at the next layer is updated by $\boldsymbol{m}^{q} = \boldsymbol{m}^{q-1} \odot \boldsymbol{m}$, where $\odot$ denotes Hadamard product. Therefore, once a feature is dropped, it will never be used in the following layers.

	\subsubsection{Differentiable Sampling}
	Although the design above is able to perform feature vector selection, it is difficult to implement during training. The main obstacle is that the sampling operation from $\boldsymbol{P}$ to obtain the selection mask vector $\boldsymbol{m}$ is non-differentiable. Note that $\boldsymbol{P}=[\boldsymbol{p}_1,\boldsymbol{p}_2,\dots,\boldsymbol{p}_{N_d} ]^{\top} \in \mathbb{R}^{N_d \times 2}$ denotes the probability matrix and the elements of the first rank and second rank are the probabilities of keeping and dropping these $N_d$ feature vectors, respectively. We take the $j$-th feature vector $\boldsymbol{f}_j^q$ in $\boldsymbol{F}^q$ as an example, and a straightforward way to determine whether to keep it is to sample from $\boldsymbol{p}_j \in \mathbb{R}^2$. However, the sampling operation is non-differentiable, which hinders the back propagation of the gradients. To address this problem, the Gumbel-Softmax technique is adopted. It is differentiable and makes it possible to train the selection module. Firstly, with the Gumbel-Max, discrete one-hot sample from the distribution, $\boldsymbol{p}_j=\big[p_j^1,p_j^2\big]$, can be expressed as
	\begin{equation}
		\boldsymbol{o} \triangleq[o_1,o_2]=\textit{oneshot}\big(\arg \max _{i=1,2} \big(\log (p_j^i)+g_i\big)\big),
	\end{equation}
	where $g_i$ for $i=1,2$ are samples drawn from $\textit{gumbel}(0, 1)$, the operation $\textit{oneshot}(n)$ denotes generating a one-hot vector, where the $n$-th element equals $1$. For instance, $\textit{oneshot}(1)=[1,0]$. The $\textit{gumbel}(0, 1)$ distribution can be sampled using inverse transform sampling by drawing $u$ from a uniform distribution $U(0,1)$ and computing $g=-\log (-\log (\mathrm{u}))$.
	
	To approximate the non-differentiable $\arg \max(\cdot)$, we further use the softmax function as a continuous, differentiable approximation \cite{gumbel}:
	\begin{equation}
		\hat{o}_i=\frac{\exp \left( \left(\log (p_j^i)+g_i\right) / \tau\right)}{\sum_{k=1}^2 \exp \left(\left(\log (p_j^k)+g_k\right) / \tau\right)}, \quad 
	\end{equation}
	for $i=1,2$, where $\tau$ is a temperature parameter that controls the discreteness. As the softmax temperature $\tau$ approaches $0$, sample from the Gumbel distribution, i.e., $\hat{\boldsymbol{o}}=\left[ \hat{o}_1, \hat{o}_2\right]$, becomes one-hot. At higher temperatures, it is no longer one-hot, and becomes uniform. It is differentiable and makes it possible to train the selection module. Therefore, the $i$-th element in selection vector $\boldsymbol{m}$ can be written as 
	\begin{equation}   
		m_i\!=\!\hat{o}_1\!=\!\textit{ gumbel-softmax }(\boldsymbol{p}_i)[1].
	\end{equation}
	In particular, the output of $\textit{ gumbel-softmax}(\cdot)$ is an approximate one-hot vector with the same shape as the input, where element $1$ indicates the sampled result. Thus, the output one-hot vector $\textit{gumbel-softmax}(\boldsymbol{p}_i)$ has two elements, and we take its first element as the result. Therefore, $\boldsymbol{m}$ will be a mask vector sampled from $\boldsymbol{F}^q$, with each element being $0$ or $1$.
	
	\begin{figure*}[t]
		\begin{centering}
			\includegraphics[width=0.59\textwidth]{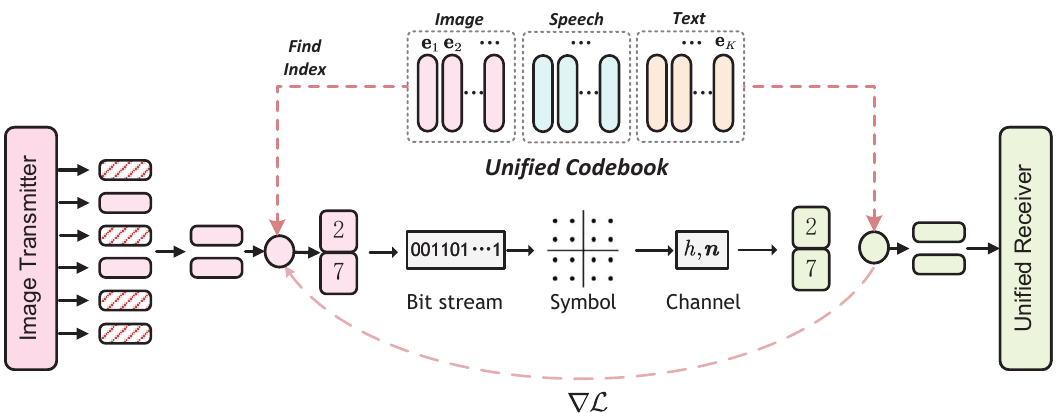}
			\par\end{centering}
		\caption{The unified codebook for digital data transmission. }
		\label{DomainAndCodebook}
	\end{figure*}

	\subsubsection{Training and Inference}
	
	The training process for U-DeepSC encompasses the dynamic channel encoder, ensuring its capability to make decisions regarding the transmission of specific feature vectors. To optimize dynamic transmission overhead across varying channel conditions, the ratio of transmitted feature vectors is governed by a predetermined variable value  $\delta = f_n\big( \sigma^2; \bm{\psi}\big)$, where $\bm{\psi}$ denotes the trainable parameters and  $\sigma^2$ denotes the variable channel noise. In practical communication systems, the receiver has the capability to gain awareness of channel conditions through channel feedback. Consequently, the encoding of features can be enhanced by integrating the knowledge of channel conditions into the process. In particular, when $f_n\big( \sigma^2; \bm{\psi} \big)$ is configured as a non-negative, monotonically increasing function implemented by DNNs, and a greater number of feature vectors will be chosen for larger $\sigma ^2$. To implement this non-negative increasing function, we employ
	a series of multi-layer perceptron (MLP) layers. The mapping of an MLP layer can be represented by
	\begin{equation}
		\textit{MLP}(\bm{x})=\textit{tanh}(\bm{W}\bm{x}),
	\end{equation}
	where we employ $\textit{tanh}(\cdot)$ as the activation function, $\bm{W}$ denotes the trainable parameters, and $\bm{x}$ represents the input vector.
	Since we aim to design $f_n$ as a non-negative increasing function with respect to the channel noise, we let $\bm{W}=\textit{abs}(\bar{\bm{W}})$ for each MLP layer, where $\bar{\bm{W}}$ is the actual trainable parameters of each MLP layer.  The monotonicity can be easily acknowledged by deriving the derivatives with respect to $\bm{x}$.

	Given the specific channel noise $\sigma ^2$, we obtain a set of target ratios for $N_q$
	corresponding selection modules, i.e., $[\delta,\delta^2,...,\delta^{N_q}]$. We apply the MSE loss to supervise the prediction module:
	\begin{equation}\label{ratioloss1}
		\mathcal{L}_{\textit{FSM}}=\frac{1}{N_q} \sum_{q=1}^{N_q}\left(\delta^{q}-\frac{1}{N_d} \sum_{i=1}^{N_d} m_i^{q}\right)^{2},
	\end{equation}
	where $\frac{1}{N_d} \sum_{i=1}^{N_d} m_i^{q}$ denotes the true ratio of transmitting feature vectors. By supervising the true ratio with the target ratio, only the target ratio of feature vectors will be processed by the decoder for training. However, if we apply loss (\ref{ratioloss1}) directly, more feature vectors will be selected as the training goes on, since the model tends to keep more feature vectors to improve the performance. To balance the performance and the number of transmitted symbols, we add the $l_1$-norm of $\delta$ to loss (\ref{ratioloss1}), as given by
	\begin{equation}\label{ratioloss2}
		\mathcal{L}_{\textit{FSM}}=\frac{1}{N_q} \sum_{q=1}^{N_q}\left(\delta^{q}-\frac{1}{N_d} \sum_{i=1}^{N_d} m_i^{q}\right)^{2} + \| \delta \|.
	\end{equation}
	It makes the model keep fewer feature vectors with the decrement of $\delta$,  and enables the model to achieve a good performance by only transmitting a part of feature vectors.

	During the inference phase, with knowledge of the channel noise variance, we can selectively discard less informative feature vectors based on probabilities generated by the selection modules. This ensures that only $N_d\delta^{N_q}$ feature vectors are transmitted to the receiver. In particular, only the top $N_d \delta^{N_q}$ feature vectors with the highest probabilities are transmitted. In addition, only the important parts of the features are retained after FSM, while the unimportant parts of feature vectors are dropped directly. Furthermore, the FSM retains only the essential  features, discarding unimportant parts of the feature vectors. As a result, the transmitter transmits only the retained feature vectors without the need to transmit the mask. At the receiver, the received feature vectors are concatenated and directly fed into the decoder.

	\section{Unified Codebook and Receiver}
	In this section, we first design the unified codebook for the multi-task applications. Then, we design the unified receiver based on Transformer decoder.
	
	\subsection{Unified Codebook for Multimodal Data}
	We aim to design a unified codebook for all considered tasks with different modalities of data to enable digital transmission. The codebook consists of a number of basis vectors and the encoded feature vectors will be represented by the basis vectors. Since the data characteristics of different modalities differ from each other, it is difficult to share the basis vectors among the data of different modalities. In contrast, the semantic information of different tasks from the same data modality may overlap. For example, the image reconstruction task and image classification task can share some basis vectors since the class semantic information must be included in global semantic information, which can be employed to reconstruct the data. It poses a potential to share the basis vectors in the  codebook among these tasks, which leads to a much smaller codebook size for multiple tasks. Therefore, we aim to design a unified codebook with three sub-codebooks for the image, text, and speech, as shown in Fig. \ref{DomainAndCodebook}(a). Different tasks of the same modality share the same sub-codebook.

	As presented in Fig. \ref{DomainAndCodebook}(a), we design a unified codebook as $\mathcal{E}\triangleq  \big\{ \mathcal{E}^v,\mathcal{E}^t,\mathcal{E}^s\bigr\} $. Specifically, 
	\begin{equation}
		\mathcal{E}^i \triangleq \big\{  \boldsymbol{e}_j^{i}\big\}_{j=1}^{M_i}, i\in\{v,t,s \},
	\end{equation}
	where $M_v$, $M_t$, and $M_s$ are the sizes of the image sub-codebook $\mathcal{E}^v$, text sub-codebook $\mathcal{E}^t$, and speech sub-codebook $\mathcal{E}^s$ respectively. $M_c=M_v+M_t+M_s$, is the total size of the unified codebook. Then, we take the image data as an example to show the way to represent the encoded feature vectors with the codebook. Recall that the input data $\boldsymbol{s}^v$ passes through the encoder to produce the encoded feature matrix $\boldsymbol{X}^v_{en}=[\bm{x}_1^v,\bm{x}_2^v,...,\bm{x}_{N_v+1}^v] \in \mathbb{R}^{(N_v+1) \times E_v}$. Then, it is represented by a group of basis vectors. Specifically, each feature vector $\boldsymbol{x}_i^v$ is represented by  $f_{vq}(\boldsymbol{x}_i^v)$, which is the nearest vector in the codebook \cite{VQVAE1},
	\begin{equation} \label{VQVAEform}
		f_{vq}(\boldsymbol{x}_i^v)=\textrm{arg} \min\limits_{ \boldsymbol{e}_{j}^v \in \mathcal{E}^v } \big\| \boldsymbol{x}_i^v - \boldsymbol{e}_{j}^v \big\|_{2}.
	\end{equation}

	The basis vectors  in the codebook are trained together with the parameters of encoder and decoder. However, the operation in \eqref{VQVAEform} is non-differentiable. Hence, the gradients are copied from the input of the decoder to the output of the encoder. In this way, the gradient is passed to the encoder to enable the back propagation. In particular, the trainable parameters of the encoder, decoder, and codebook are updated via the loss function below
	\begin{equation} \label{VQVAEloss}
		\mathcal{L}_{c}(\boldsymbol{x}_i^v;  \boldsymbol{e}_{j}^v) = \big\| \textrm{ng}\big[ \boldsymbol{x}_i^v\big] - \boldsymbol{e}_{j}^v \big\|_{2}^{2}+  \beta \big\| \boldsymbol{x}_i^v - \textrm{ng}\big[ \boldsymbol{e}_{j}^v \big] \big\|_{2}^{2},
	\end{equation} 
	where  $\beta$ denotes the hyper-parameter, and symbol $\textrm{ng}\big[\boldsymbol{u}_i^v\big]$ represents stop-gradient operator that has zero gradient during backward propagation and is the identity layer during the forward propagation. Additionally, if the task involves more than one modality, we need to calculate the sum of codebook losses of involved modalities.  Since we use the straight gradient estimation of mapping from $\boldsymbol{x}_i^v$ to $	f_{vq}(\boldsymbol{x}_i^v)$, the basis vectors  $\{\mathbf{e}_{j}^v, \forall j\}$ have no gradients from the loss function. Therefore,  to learn the basis vectors, we employ the $l_{2}$ error to move the basis vectors towards the encoded features, $\boldsymbol{x}_i^v$, as shown in the first term of \eqref{VQVAEloss}. Since the volume of the encoded feature space is dimensionless, the codebook can grow arbitrarily and cause the training process to diverge if the basis vectors, $\{\mathbf{e}_{j}^v, \forall j\}$, are not trained as fast as the encoder parameters \cite{VQVAE1}. To address this issue, we add the second term in \eqref{VQVAEloss}. In summary,  the basis vectors are optimized by the first loss term, and the encoder is optimized by the second loss term. These details have been elucidated more explicitly in the revised manuscript.

	\subsection{Unified Receiver Design}
	\subsubsection{Transformer-Based Decoding}
\begin{figure}[t]
		\begin{centering}
			\includegraphics[width=0.49\textwidth]{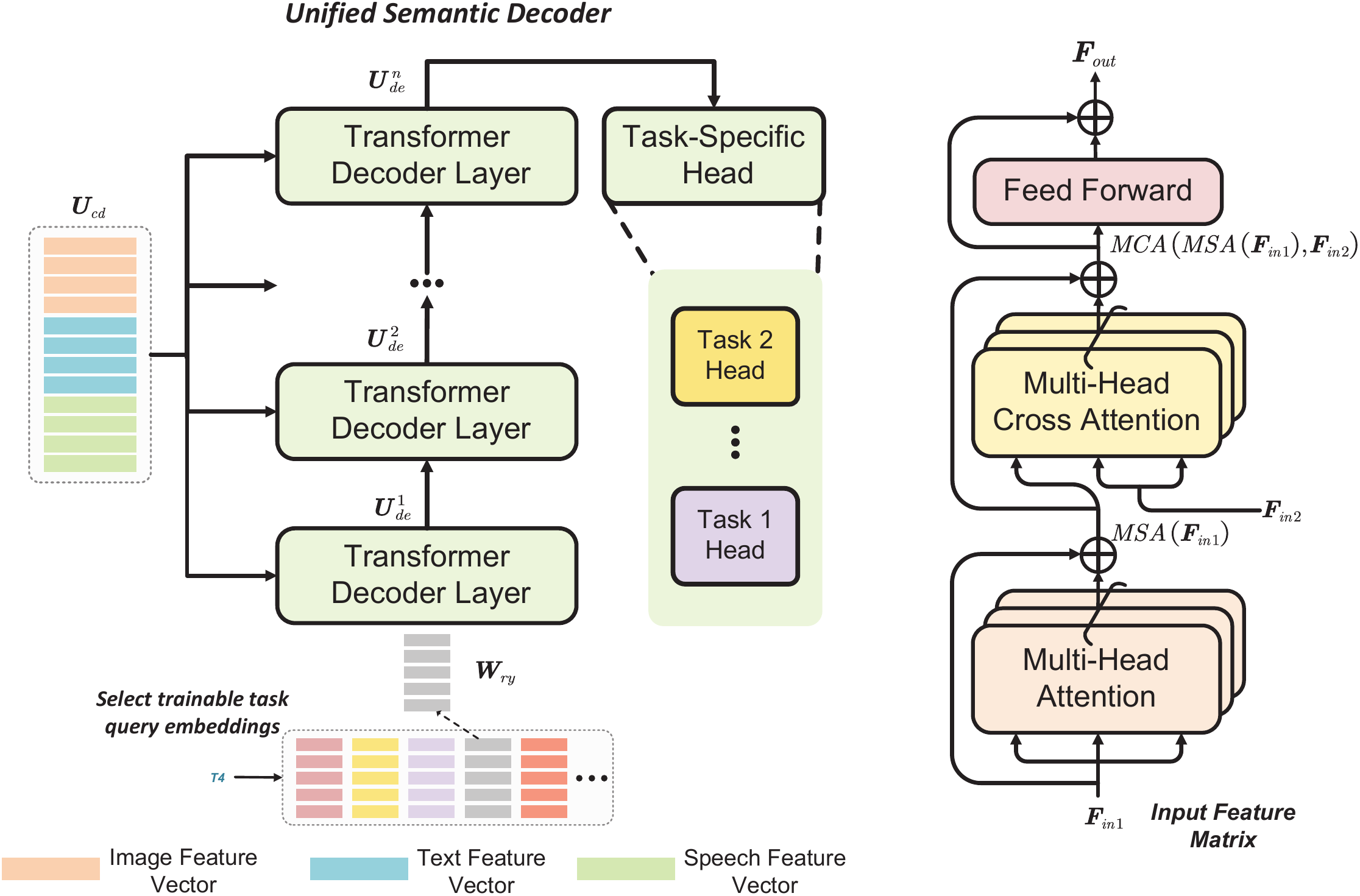}
			\par\end{centering}
		\caption{The receiver architecture of the proposed U-DeepSC.}
		\label{UDecoder}
	\end{figure}

	As depicted in Fig. \ref{UDecoder}, the Transformer decoder primarily consists of two components: MSA and multi-head cross-attention (MCA). In contrast to MSA, MCA focuses on attending to the extra input to capture relevant information. By incorporating both self-attention and cross-attention mechanisms, the Transformer decoder layer can adaptively  process the received information.
	In particular, 
	the decoding procedure of a Transformer can be denoted as 
	\begin{equation}
		\bm{F}_{out} = \textit{MSA}(\bm{F}_{in{1}})+\textit{FFN}(\textit{MCA}(\textit{MSA}(\bm{F}_{in{1}}),\bm{F}_{in{2}})),
	\end{equation}
	where $\bm{F}_{out} \in \mathbb{R}^{L_1 \times D}$ denotes the output matrix, $\bm{F}_{in{1}} \in \mathbb{R}^{L_1 \times D}$ and $\mathbf{F}_{in{2}} \in \mathbb{R}^{L_2 \times D}$ are input matrices.
	The operation $\textit{MCA}(\cdot)$ can be described as follows.
	Firstly, for simplicity, we let $\bm{F}_1=\textit{MSA}(\bm{F}_{in{1}})$ and $\bm{F}_2=\bm{F}_{in2}$.
	Then, similar to MSA, the operation $\textit{MCA}(\cdot)$ generally takes two matrices, $\bm{F}_1 \in \mathbb{R}^{L_1 \times D}$ and $\mathbf{F}_2 \in \mathbb{R}^{L_2 \times D}$, as input, and the procedure can be expressed as
	\begin{equation}
		\begin{split}
			& \textit{MCA}(\bm{F}_1,\bm{F}_2)= \\
			&\bm{F}_{1}+\textit{concat}\underbrace{\left(\textit{CA}(\bm{F}_1,\bm{F}_2),\textit{CA}(\bm{F}_1,\bm{F}_2),...,\textit{CA}(\bm{F}_1,\bm{F}_2)\right)}_{N_s} \bm{W}.
		\end{split}
	\end{equation}
	Specifically, the cross-attention operation is given by 
	\begin{equation}\label{attention}
		\textit{CA}(\bm{F}_1,\bm{F}_2) =\textit{softmax}\left( \frac{ \bm{Q} \bm{K}^{\top}}{\sqrt{D}} \right)\bm{V},
	\end{equation}
	where $\bm{Q} \in \mathbb{R}^{L_1 \times D_s}$, $\bm{K} \in \mathbb{R}^{L_2 \times D_a}$, and $\bm{V} \in \mathbb{R}^{L_2 \times D_a}$ are generated from three linear projections as 
	\begin{equation}
		\bm{Q}=\bm{F}_{1}\bm{W}_Q,\quad\bm{K}=\bm{F}_{2}\bm{W}_K, \quad\bm{V}=\bm{F}_{2}\bm{W}_V,
	\end{equation}
	where $\bm{W}_Q\in \mathbb{R}^{D \times D_a}$, $\bm{W}_K\in \mathbb{R}^{D \times D_a}$, and $\bm{W}_V \in \mathbb{R}^{D \times D_a}$. 
	By connecting multiple Transformer decoder layer, a Transformer decoder is obtained.

	\subsubsection{Unified Semantic Decoder}
	The received encoded features are firstly processed by the channel decoder, whose output is denoted as $\boldsymbol{U}_{cd}$, as shown in Fig. \ref{UDecoder}. For image-only tasks, text-only tasks, and speech-only tasks, the input to the decoder can be represented by $\boldsymbol{U}_{cd}=\boldsymbol{U}^{v}$, $\boldsymbol{U}_{cd}=\boldsymbol{U}^{t}$, and $\boldsymbol{U}_{cd}=\boldsymbol{U}^{s}$,  respectively, where $\boldsymbol{U}^{v}$, $\boldsymbol{U}^{t}$, and $\boldsymbol{U}^{s}$ denote the decoded image, text, and speech feature matrices processed by channel decoder, respectively. For multimodal tasks, we concatenate the decoded features from the corresponding modalities of data into a sequence, e.g., $\boldsymbol{U}_{cd}= \textit{concat}\left(\boldsymbol{U}^{v}, \boldsymbol{U}^{t}\right)$ for image-and-text tasks.
	
	Unlike the separate design for each modality at the transmitter, the semantic decoder is built upon the unified Transformer decoder structure. As shown in Fig. \ref{UDecoder}, the semantic decoder takes the output of channel decoder, $\boldsymbol{U}_{cd} \in \mathbb{R}^{N_p \times E_p} $, and the task-specific query  matrix, $\boldsymbol{W}_{ry}\in \mathbb{R}^{N_r \times E_p}$, as the input, where $N_p$ is the total number of feature vectors, $N_r$ and $E_p$ denote the number and length of task-specific query embedding vectors, respectively. The task-specific query embedding matrix serves as an indicator of the task assigned to the semantic decoder, akin to the conventional input required by the Transformer decoder \cite{Bert}. Similar to the task embedding vectors, we would generate several task-specific query embedding matrices, and each of which corresponds to one task. During the testing phase, only the task-specific query embedding matrix relevant to the ongoing task is utilized.  To illustrate, if there are a total of $6$ tasks, we create $6$ trainable task-specific query embedding matrices. In the inference phase, we select the corresponding matrix as input to indicate the intended task, as illustrated in Fig. \ref{UDecoder}. Let $\bm{F}_{in1}=\boldsymbol{W}_{ry}$ and $\bm{F}_{in2}=\boldsymbol{U}_{cd}$. Then, we obtain the output of the $i$-th decoder layer $\boldsymbol{U}^{i}_{de}$. Therefore, the output of the first decoder layer $\boldsymbol{U}^{1}_{de}$ can be denoted by
	\begin{equation}
		\boldsymbol{U}^{1}_{de} = \textit{MSA}(\boldsymbol{W}_{ry})+\textit{FFN}(\textit{MCA}(\textit{MSA}(\boldsymbol{W}_{ry}),\boldsymbol{U}_{cd})).
	\end{equation}
	Then, for the later layers, the output of the $i$-th decoder layer $\boldsymbol{U}_{i}^{de}$ can be iteratively denoted as 
	\begin{equation}
		\boldsymbol{U}^{i}_{de} = \textit{MSA}(\boldsymbol{U}^{i-1}_{de})+\textit{FFN}(\textit{MCA}(\textit{MSA}(\boldsymbol{U}^{i-1}_{de}),\boldsymbol{U}_{cd})).
	\end{equation}

	\section{Masked Cross-Attention and Joint Training}
	In this section, we design the masked cross-attention for parallel training. The overall training algorithm is also introduced.
	\subsection{Masked Cross-Attention for Parallel Training}
	Recall that FSM can select informative feature vectors using Gumbel-Softmax sampling. However, the sampling operation is inherently random, leading to an indeterminate number of selected features that can vary across different inputs. Consequently, the received feature matrices $\mathbf{U}_{cd}$, which consist of selected feature vectors for distinct inputs (such as different images), exhibit varying shapes. The inherent variability in feature vector sizes poses a challenge to achieving parallel batch-wise training, as conventional deep learning frameworks often demand uniform feature shapes within a batch. Diverging from the approach presented in \cite{MultiResolution}, which utilizes multiple decoders to handle varying feature vector sizes, our solution for efficient parallel training involves the introduction of a masked cross-attention mechanism.

	To this end, we do not directly drop the redundant feature vectors, instead, all feature vectors are kept in the training phase and we only drop the redundant feature vectors in the inference phase feature. Thus, we must keep the number of feature vectors unchanged in the training phase, while cutting down the interactions between the redundant feature vectors and other informative feature vectors. Nevertheless, merely zero-out the redundant feature vectors according to the mask vector is not feasible, since the zeroed feature vectors will still influence other vectors through the Softmax operation, as given in (\ref{attention}). In order to drop the redundant feature vectors, we mask the attention scores calculated by the redundant feature vectors. That is, the main procedure of the masked cross-attention is described as follows. We first denote the attention score matrix as
	\begin{equation}
		\bm{A}=\bm{Q K}^T / \sqrt{D} \in \mathbb{R}^{L_1 \times L_2}. \\
	\end{equation}
	Then, we generate a mask matrix $\bm{M} \in \mathbb{R}^{L_1\times L_2}$ at the decoder for $\bm{A}$ based on the final mask vector $\hat{\bm{m}} \in \mathbb{R}^{L_2}$:
	\begin{equation}
		\bm{M}[i]= \hat{\bm{m}}, \;1 \leq i  \leq L_1.
	\end{equation}
	Accordingly, we denote the cross-attention matrix in (\ref{attention}) as $\bm{B}$, which can be calculated by 
	\begin{equation}
		\bm{B}[i,j]=\frac{\exp \left(\bm{A}[i,j]\right) \bm{M}[i,j]}{\sum_{l=1}^{L_2} \exp \left(\bm{A}[i,l]\right) \bm{M}[i,l]}.
	\end{equation}
	Particularly, $\bm{M}[i,j]=1$ indicates that the $j$-th feature vector will influence the $i$-th feature vector. In this way, the redundant feature vectors will not contribute to the informative feature vectors, so that they can be directly dropped in the inference phase. 
	
		\begin{algorithm}[t] 
		\caption{Two-phase training algorithm} 
		\label{Training}
		\SetAlgoLined
		\textbf{Input:} Training datasets consist of input image, text, speech data and labels. The numbers of training epochs for two phases, $N_1$ and $N_2$, the learning rate. \\
		\textbf{Output:} Optimized parameters $\{\bm{\theta}_v^*,\bm{\theta}_s^*,\bm{\theta}_t^*,\bm{\phi}_{u}^*\}$. \\
		\textbf{First Phase:}\\
		Fix parameters of $f_n$. \\
		\For{$i\leftarrow 1$ \KwTo $N_1$}{
			Choose one task and generate a batch of samples. \\
			Generate the selection mask vectors.\\
			Compute FSM loss $\mathcal{L}_{\textit{FSM}}$ based on (\ref{ratioloss1}).\\
			Continue forward propagation with the generated mask vectors.\\
			Compute the task-specific loss $\mathcal{L}_{p}$ according to the type of task.\\
			Compute the total loss $\mathcal{L}\left(\bm{s}^v,\bm{s}^t,\bm{s}^s\right)=\mathcal{L}_{p}+\mathcal{L}_{\textit{FSM}}$.\\
			Update parameters of U-DeepSC, $\{\bm{\theta}_v,\bm{\theta}_s,\bm{\theta}_t,\bm{\phi}_{u}\}$, using $\mathcal{L}\left(\bm{s}^v,\bm{s}^t,\bm{s}^s\right)$.\\
		} 
		\textbf{Second Phase:}\\
		Load the parameters trained in the first phase. \\
		\For{$i\leftarrow 1$ \KwTo $N_2$}{
			Choose one task and generate a batch of samples. \\
			Sample the channel variance $\sigma^{2}$ from the given SNR range.\\
			Sample the channel gain coefficient $h$.\\
			Generate the selection mask vectors before transmitting.\\
			Compute FSM loss $\mathcal{L}_{\textit{FSM}}$ based on (\ref{ratioloss2}).\\
			Compute codebook loss $\mathcal{L}_{c}$ based on (\ref{VQVAEloss}).\\
			Continue forward propagation with the generated mask vectors.\\
			Compute the task-specific loss $\mathcal{L}_{p}$ according to the type of task.\\
			Compute the total loss $\mathcal{L}\left(\bm{s}^v,\bm{s}^t,\bm{s}^s\right)=\mathcal{L}_{p}+\mathcal{L}_{\textit{FSM}}+\mathcal{L}_c$.\\
			Update parameters of U-DeepSC, $\{\bm{\theta}_v,\bm{\theta}_s,\bm{\theta}_t,\bm{\phi}_{u},\bm{\psi}\}$, using $\mathcal{L}\left(\bm{s}^v,\bm{s}^t,\bm{s}^s\right)$.\\
		} 
	\end{algorithm}

	\subsection{Joint Training Algorithm}
	To jointly learn the considered tasks, we propose an efficient method to train the modules in the U-DeepSC system, which can be divided into two phases. 
	\begin{itemize}
		\item 
		
		In the first phase, we fix parameters of $f_n$, $\bm{\psi}$, and jointly train the other parameters. Specifically, we randomly choose a task and sample data from the corresponding dataset. After determining the involved modalities, we activate the required modules and update the parameters of encoders and decoders using the loss $\mathcal{L}\left(\bm{s}^v,\bm{s}^t,\bm{s}^s\right)$. In addition, the codebook for feature vector representation is not included in this phase. After convergence, the model is generally capable of extracting task-specific semantic information and achieving satisfactory task performance.

		\item 
		In the second phase, we fine-tune the entire U-DeepSC model, building upon the parameter trained in the first phase. Furthermore, the codebook loss $\mathcal{L}_c$ and the FSM loss $\mathcal{L}_\textit{FSM}$ are both incorporated into the overall loss function $\mathcal{L}\left(\bm{s}^v,\bm{s}^t,\bm{s}^s\right)$. This phase aims to optimize the complete system, aiming for global optimization. Consequently, a balanced trade-off between transmission overhead and task performance can be achieved for different tasks.

	\end{itemize}
	
	The detailed training procedures are summarized in Algorithm \ref{Training}. The proposed U-DeepSC is a general framework, and can support various tasks of different modalities of data, e.g., object detection and speech classification. Firstly, it necessitates the design of a task embedding vector and task-specific query embedding matrices customized to the specific characteristics of the new task. Secondly, the creation of a task-specific head dedicated to the new task is essential. Finally, the model must undergo joint learning to effectively support the new task.

	\section{Simulation Results} \label{Simulation}
	In this section, we demonstrate the superiority of the proposed U-DeepSC by numerical results.

	\subsection{Simulation Setup}
	
	The setting of the training procedure is: the AdamW optimizer with learning rate $1\times10^{-5}$, batch size $50$,  weight decay $5\times10^{-3}$\footnote{The code is available at \textit{github.com/zhang-guangyi/t-udeepsc}.}, where we find that the smaller learning rate and the larger batch size lead to better performance.  The number of FSM is $N_q=2$. 
	To verify the effectiveness of U-DeepSC, we test our U-DeepSC on the aforementioned six tasks, each corresponding to a dataset. Specifically, the CIFAR-10 dataset is utilized for both image classification and image reconstruction tasks. For text-related tasks such as text reconstruction and text classification, the SST-2 dataset is employed. Additionally, the VQAv2 dataset is chosen for the visual question answering task, and the MOSEI dataset is utilized for the video task. In the context of image reconstruction, a patch size of $4$ is set, while for image classification, it is set to $32$. The text-related tasks utilize BERT-base-uncased as the text embedding tool. For video data's vision stream, Facet, an analytical tool based on the facial action coding system (FACS), is used to extract facial features. Speech features are extracted using COVAREP, a professional acoustic analysis framework. The image and text transmitters are initialized using a pretrained vision transformer and the BERT model, respectively. Furthermore, in the simulation, our focus is on AWGN and Rayleigh fading channels.
	
	\begin{figure*}[t]
		
		\begin{centering}
			\subfloat[Text Classification]{\label{fig:a}\includegraphics[width=4.5cm]{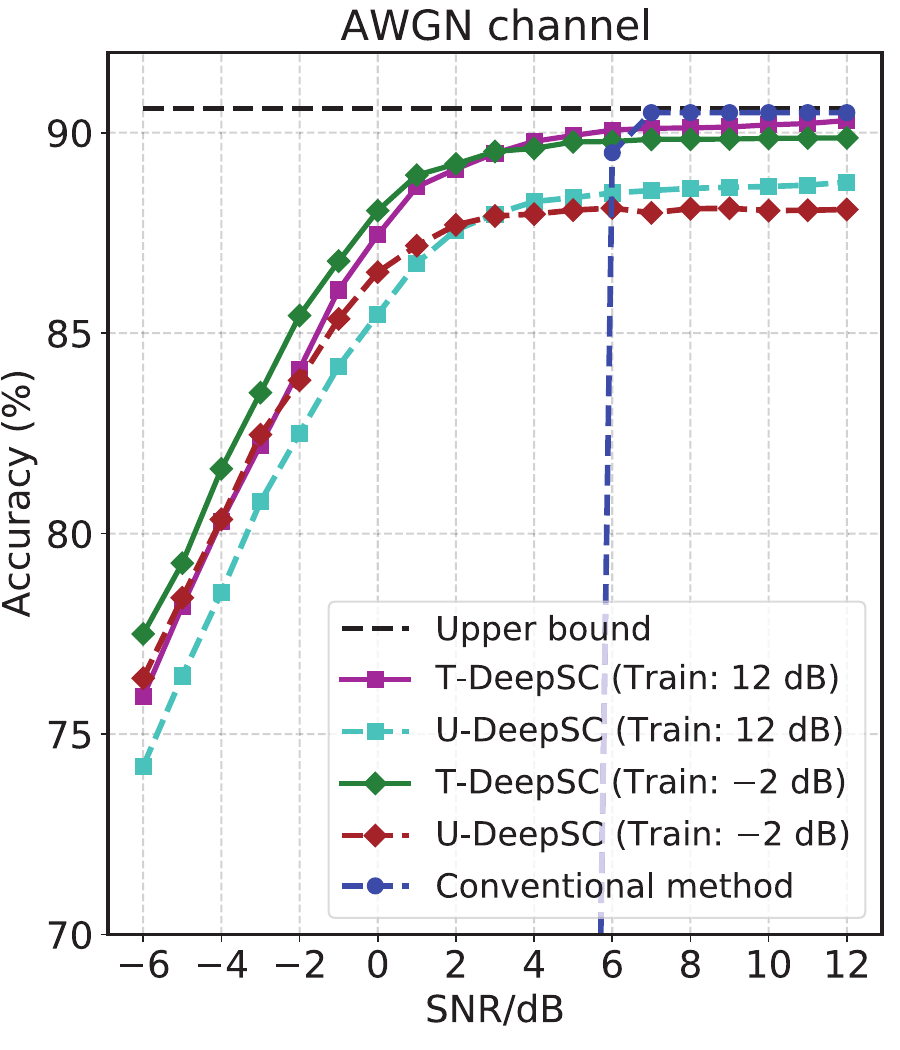}}
			\subfloat[Video Sentiment Analysis]{\label{fig:b}\includegraphics[width=4.5cm]{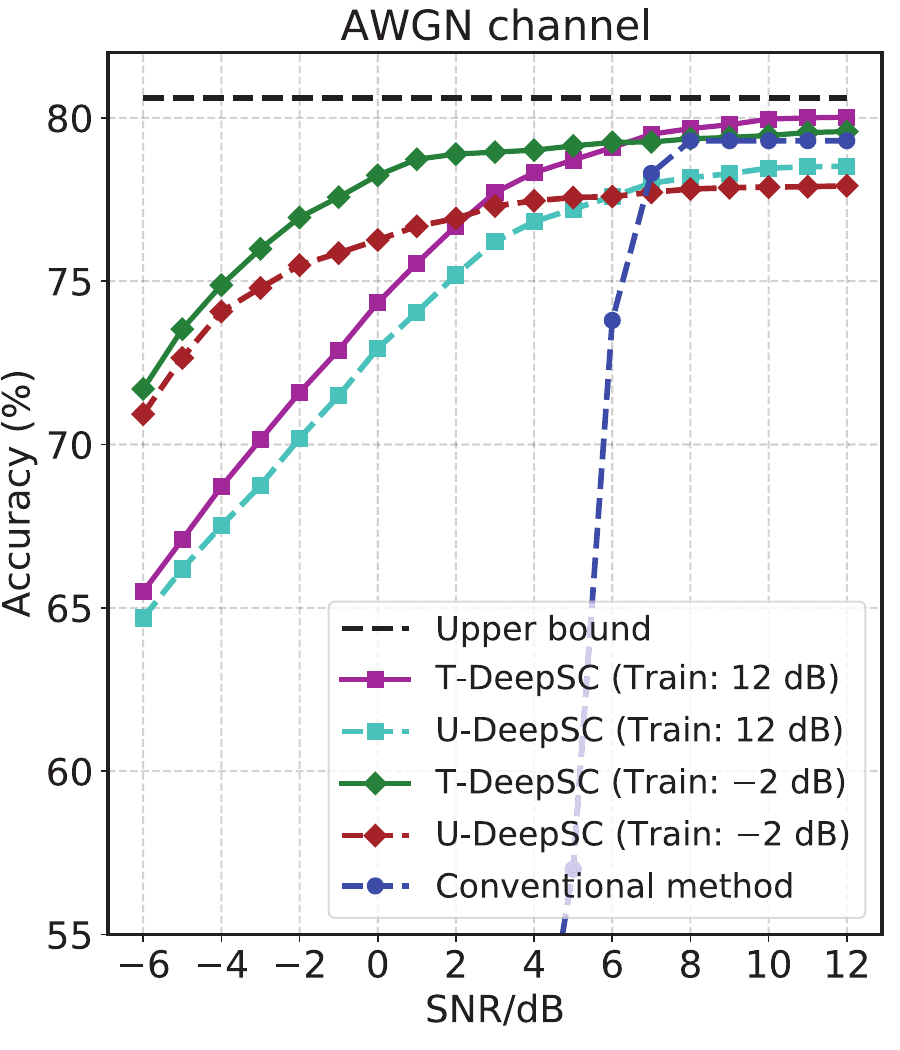}}
			\subfloat[Image Classification]{\label{fig:a}\includegraphics[width=4.5cm]{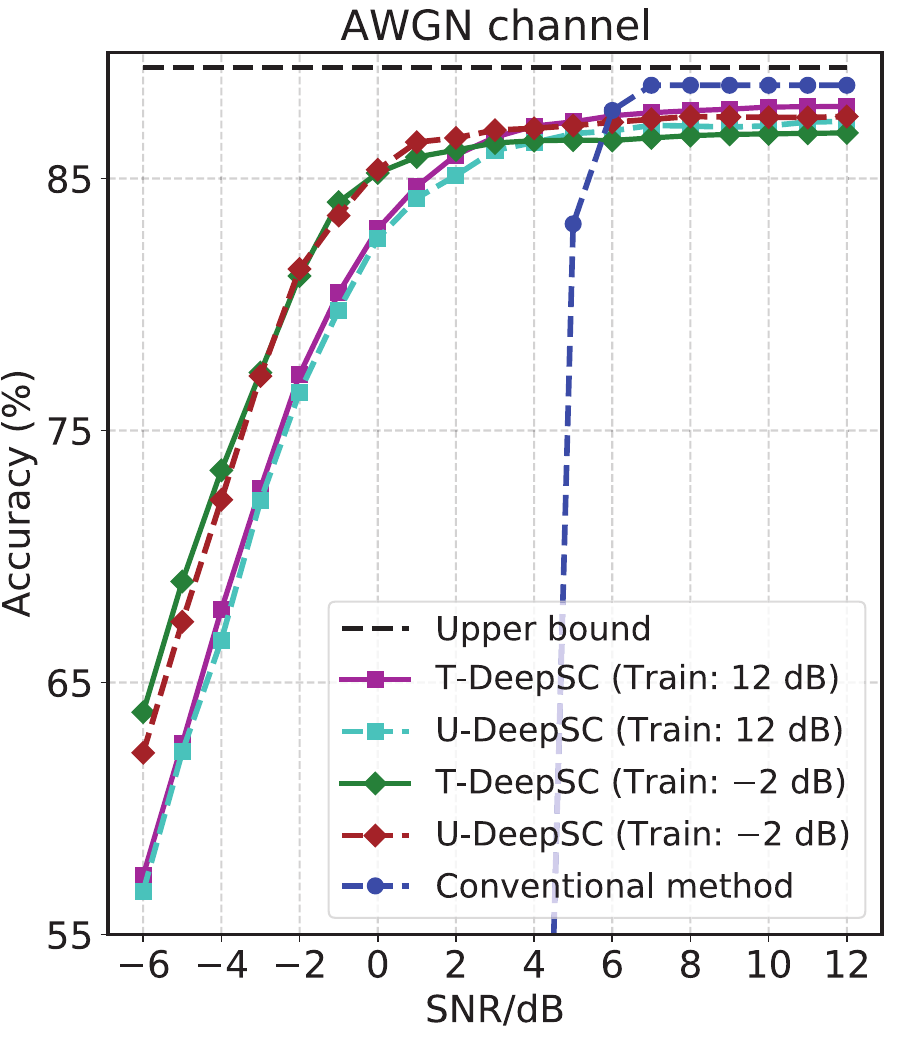}}\quad
			\subfloat[Text Reconstruction]{\label{fig:b}\includegraphics[width=4.5cm]{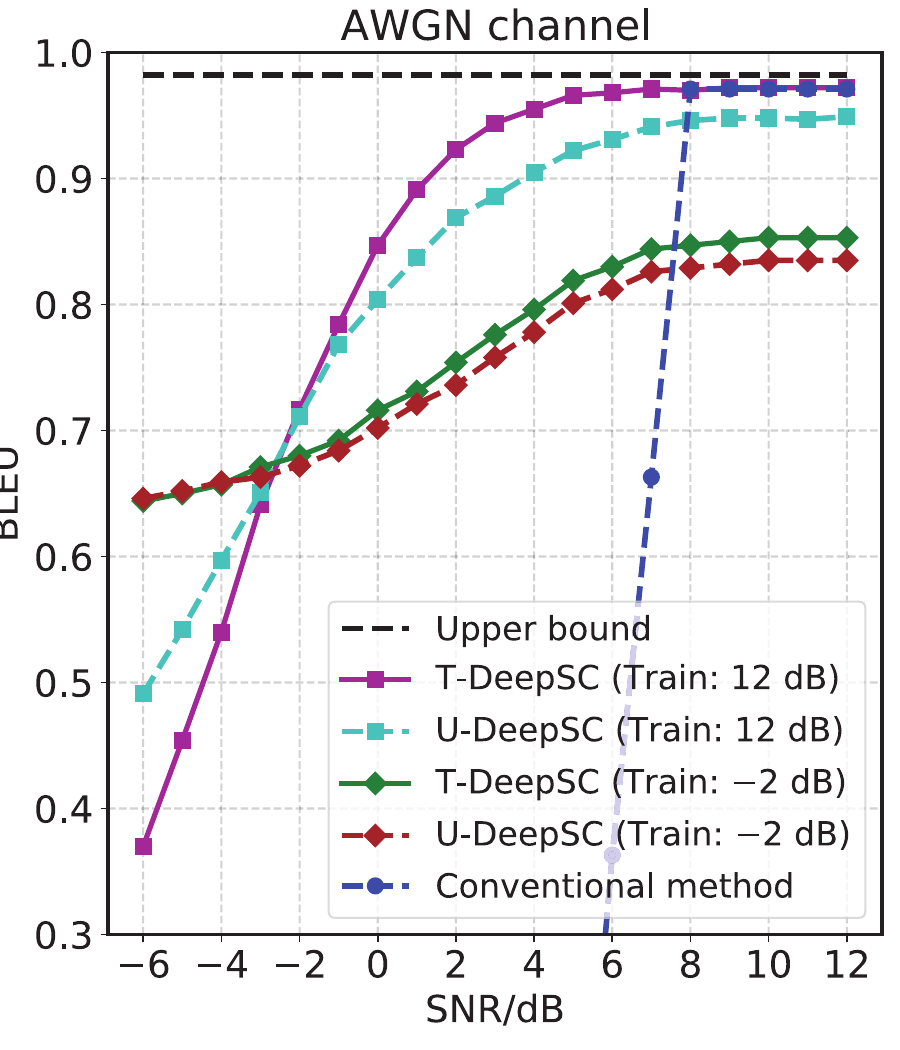}}
			\subfloat[Image Reconstruction]{\label{fig:b}\includegraphics[width=4.5cm]{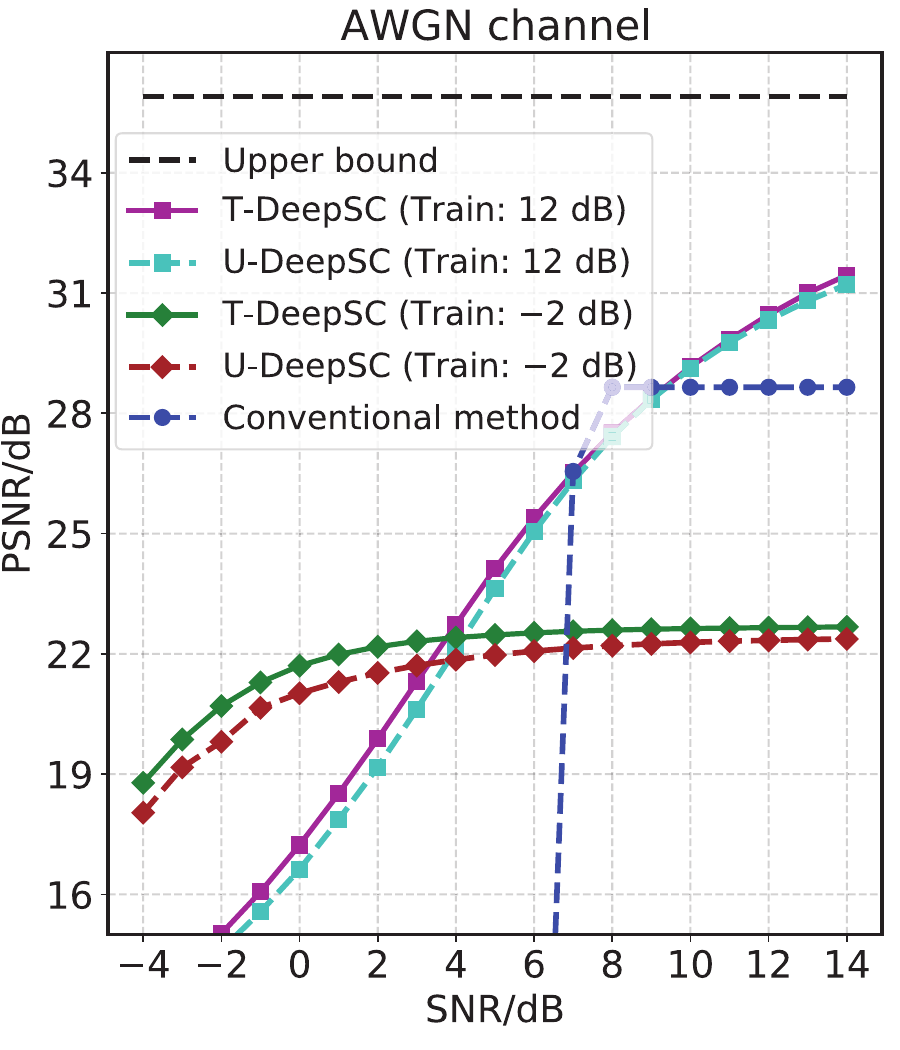}}
			\subfloat[Visual Question Answering]{\label{fig:b}\includegraphics[width=4.5cm]{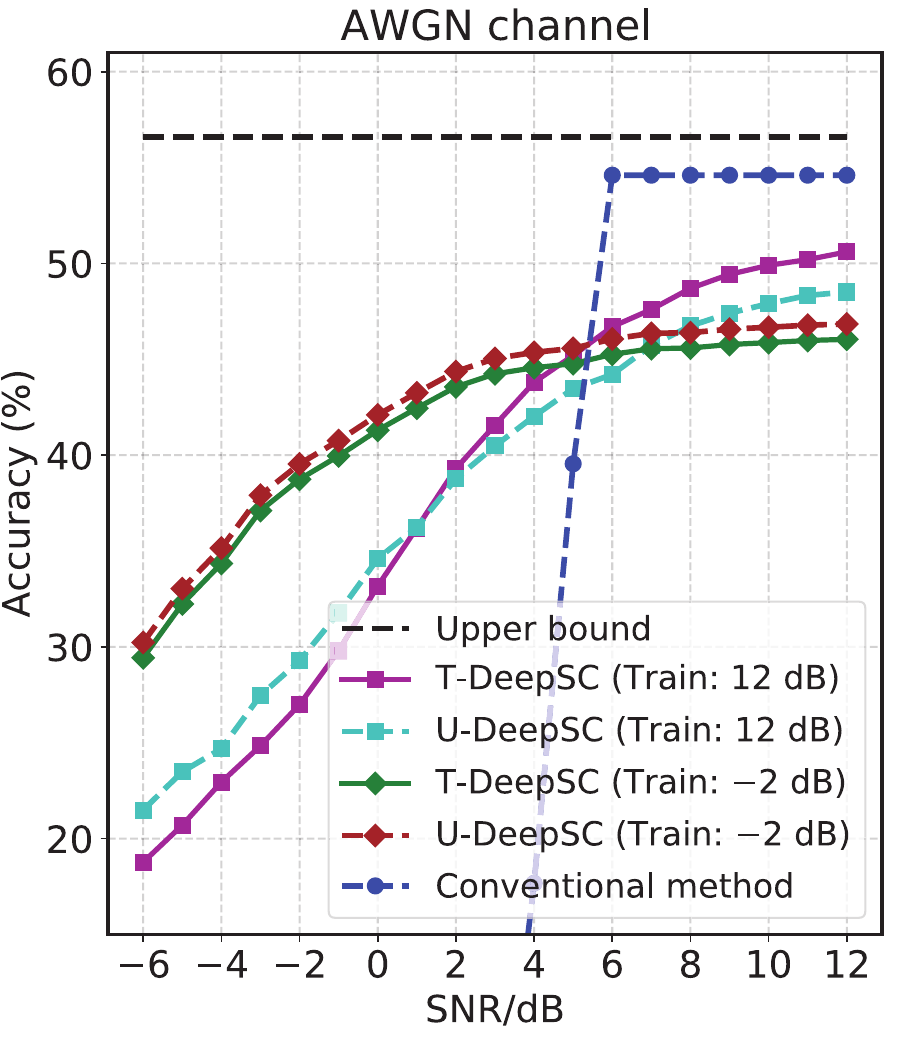}}
			
			\caption{The performance of six tasks versus SNR, including text classification, video sentiment analysis, image classification, text classification, image reconstruction, and visual question answering. T-DeepSC and U-DeepSC are both trained at SNR $=-2$ dB and SNR $=12$ dB under AWGN channels.} 
			\label{results}
		\end{centering}
	\end{figure*}

	For comparison, three benchmarks are considered. 
	\begin{itemize}
		\item Conventional methods: This is the conventional separate source-channel coding. For the image data, the joint photographic experts group (JPEG) and low-density parity-check code (LDPC) are adopted as image source coding and image channel coding, respectively. In addition, for video, we adopt the H.264 video compression codecs for source coding. For the text data, the 8-bit unicode transformation format  (UTF-8) encoding and the Turbo coding are adopted as the text source coding and text channel coding, respectively. For the speech signal, 16-bits pulse code modulation (PCM) and LDPC are employed as the source coding and channel coding, respectively. Moreover, the coding rate of channel coding is selected as $\nicefrac{1}{2}$.
		
		\item T-DeepSC: The task-oriented deep learning enabled semantic communication (T-DeepSC) designed for a specific task with the same architecture as U-DeepSC and is implemented by separately trained U-DeepSC.
		
		\item Upper bound: Results obtained via delivering noiseless image, speech, and text features to the receiver based on the T-DeepSC.
	\end{itemize} 
	
	\begin{table}[!htp]\footnotesize
		\caption{The number of transmitted symbols.}
		\centering
		\begin{tabular}{c|c}
			\toprule[0.3mm]
			Task    & Number of  symbols         \\ 
			\midrule[0.18mm]
			\midrule[0.18mm]
			Image  Classification   & 8   \\ 
			Image Reconstruction       & 8 (each patch) \\ 
			Text Classification         & 4     \\ 
			Text Reconstruction     & 12   (each word)     \\ 
			Visual Question Answering        & 30     \\ 
			Video Sentiment Analysis        & 19    \\ 
			\bottomrule[0.3mm]
		\end{tabular}
		\label{table_number}
	\end{table}

	\subsection{Task Performance}
	
	Fig. \ref{results} illustrates the performance of the investigated schemes versus the SNR for different tasks. T-DeepSC and U-DeepSC are both trained at the same SNR, i.e., $-2$ dB and $12$ dB. The average number of transmission symbols of U-DeepSC equals that of T-DeepSC. Specifically, the numbers of transmitted symbols for these tasks are shown in Table \ref{table_number}. The proposed U-DeepSC and T-DeepSC are tested in SNR from $-6$ dB to $12$ dB (We conduct testing on the image reconstruction task starting from $-4$ dB, as $-6$ dB is deemed too low for an image compression task that demands a higher channel capacity.). It is readily seen that both the U-DeepSC and T-DeepSC outperform the conventional schemes and the U-DeepSC approaches the upper bound at high SNR. Moreover, the proposed U-DeepSC achieves  close performance to the T-DeepSC in all considered tasks. Therefore, our proposed U-DeepSC is able to simultaneously handle six tasks with comparable performance to the task-oriented models designed for a specific task. Given that UDeepSC transmits only a specific subset of the overall features for different tasks, its satisfactory performance underscores the ability of U-DeepSC to effectively capture and specify task-specific semantic information.  By comparing the results of training at $-2$ dB with that of training at $12$ dB, we observe that the performance of U-DeepSC does not saturate immediately with improvements in channel conditions. This characteristic enhances the robustness of U-DeepSC to fluctuations in channel quality, proving advantageous, especially in scenarios involving transmission over time-varying channels.

\begin{figure}[t]
		\begin{centering}
			\subfloat[]{\label{overhead1}\includegraphics[width=4.5cm]{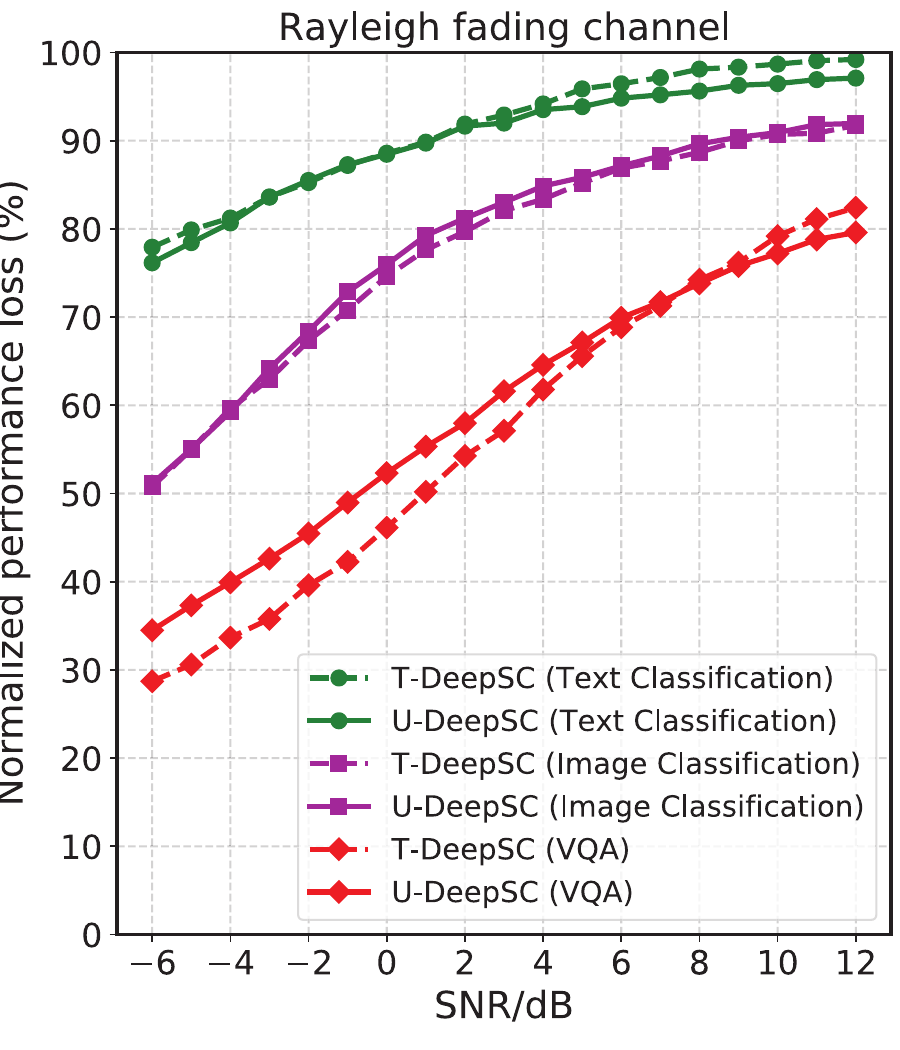}}
			\subfloat[]{\label{overhead2}\includegraphics[width=4.5cm]{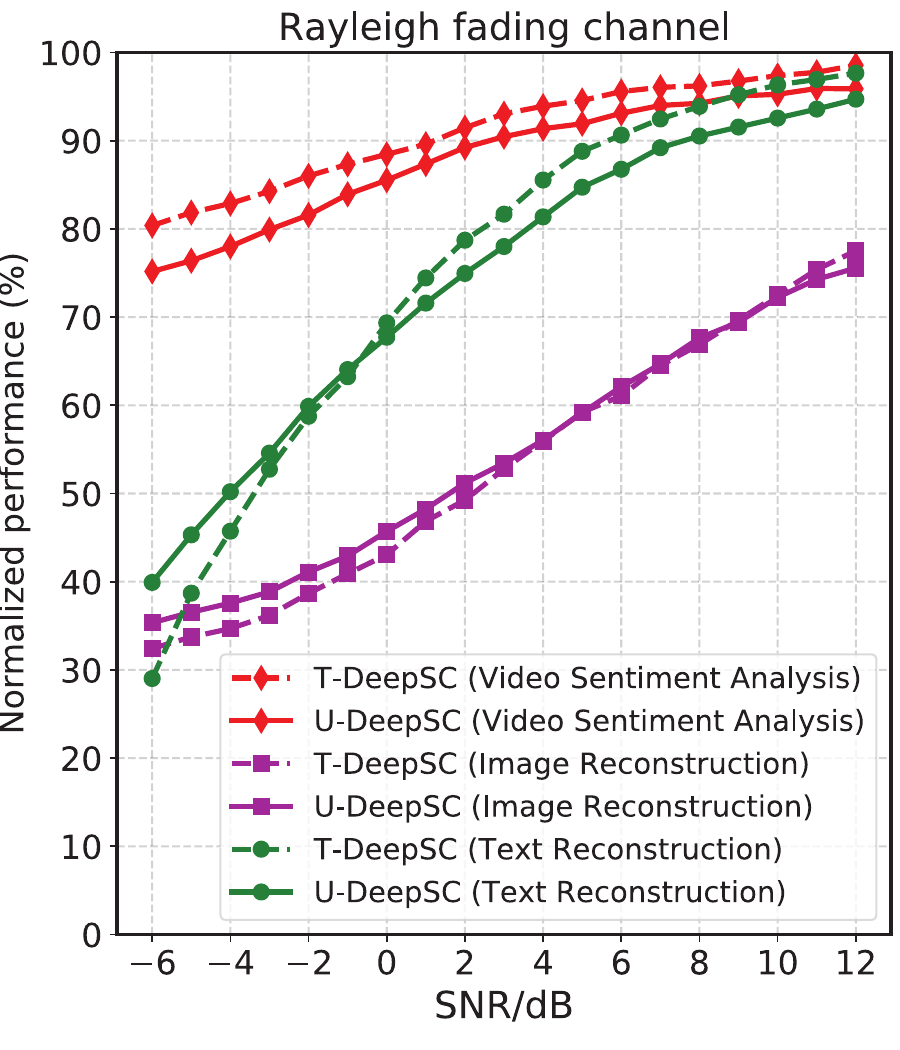}}\quad
			\caption{The performance of six tasks versus SNR under Rayleigh fading channel.} 
			\label{Rayleigh}
		\end{centering}
	\end{figure}

	We then train the proposed U-DeepSC model in the AWGN channel with SNR $=12$ dB and test it in the Rayleigh fading channel for SNR from $-6$ dB to $12$ dB. The channel information is both known at the transmitter and receiver. To show results in the same figure, we calculate the normalized performance of these tasks as the performance metric, which is obtained by dividing the actual performance by the performance achieved under error free transmission, i.e., upper bound. The results of the considered tasks are shown in Fig. \ref{Rayleigh}.  As we can see, the fading effect would lead to performance loss and the performance of U-DeepSC increases with SNR. It can be also found that the proposed model trained in AWGN could well generalize to different channels with different system settings. Moreover, the performance gap between U-DeepSC and T-DeepSC does not decrease a lot as the channel distribution changes. We further find that when we jointly learn multiple tasks, the model tends to perform better at low SNR regimes. This is mainly because the training of the other tasks acts as a perturbation, which is similar to training the model with the lower SNR.

	\begin{figure}[t]
		\begin{centering}
			\subfloat[]{\label{overhead1}\includegraphics[width=4.3cm]{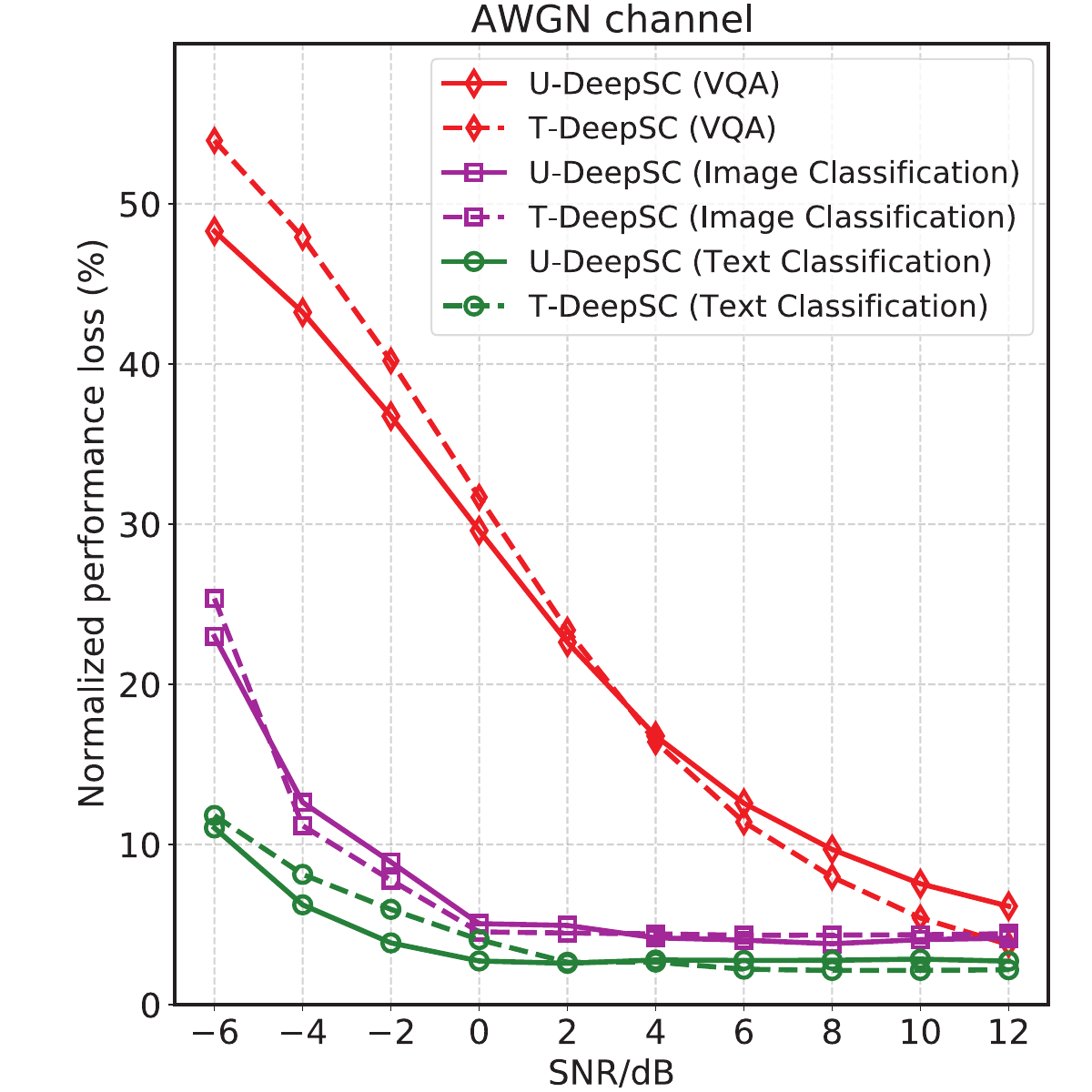}}
			\subfloat[]{\label{overhead2}\includegraphics[width=4.3cm]{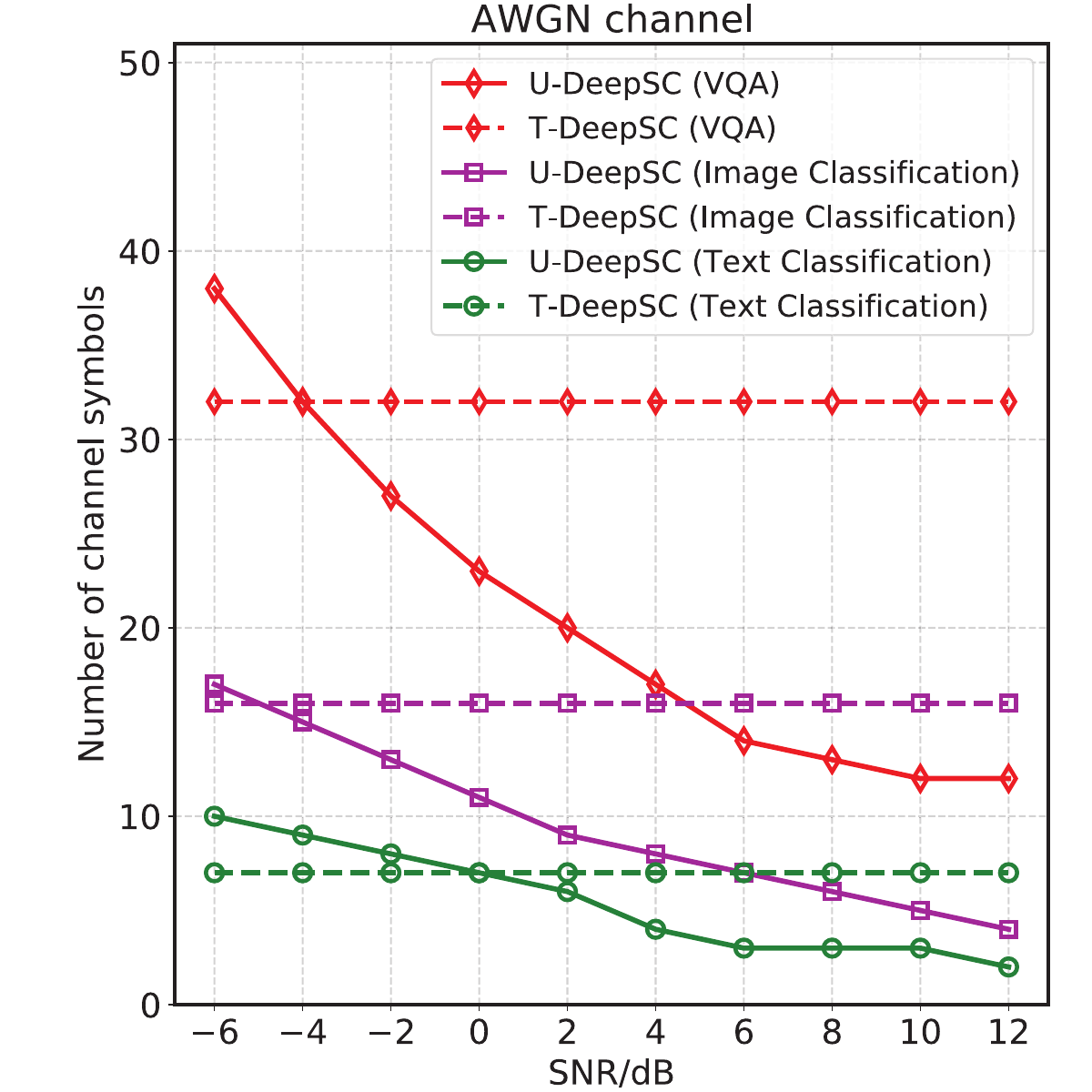}}\quad
			\caption{Normalized performance loss and number of transmitted symbols as a function of the channel SNR in dynamic channel conditions.} 
			\label{overhead12}
		\end{centering}
	\end{figure}
	
	\begin{figure}[t]
		\begin{centering}
			\subfloat[Image]{\label{overhead5}\includegraphics[width=4.4cm]{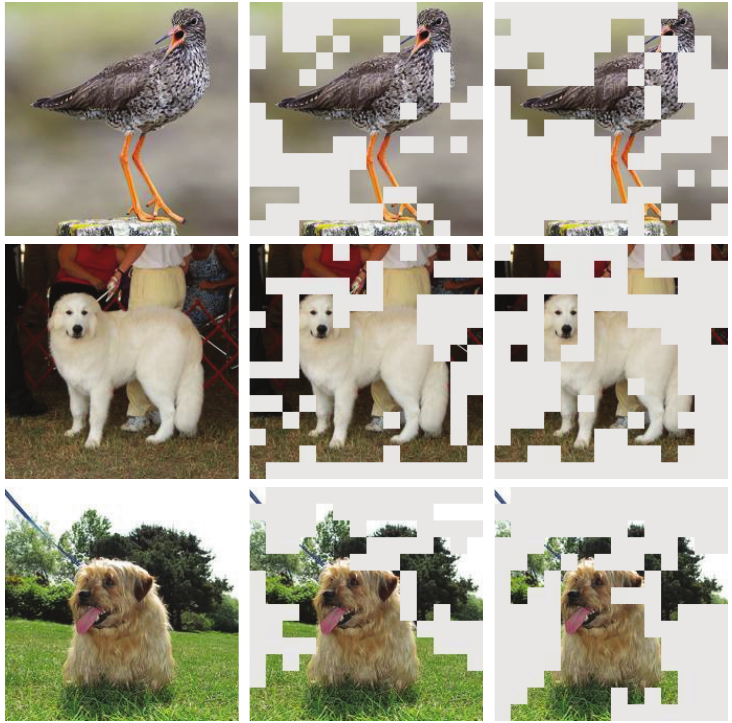}}\quad
			\subfloat[Text]{\label{overhead6}\includegraphics[width=6cm]{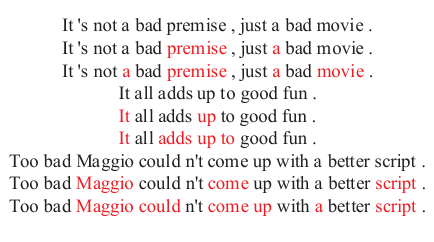}}\quad
			
			\caption{The visualization of the feature selection.} 
			\label{overhead56}
		\end{centering}
	\end{figure}

	\begin{figure}[t]
		\begin{centering}
			\subfloat[]{\label{overhead1}\includegraphics[width=4.5cm]{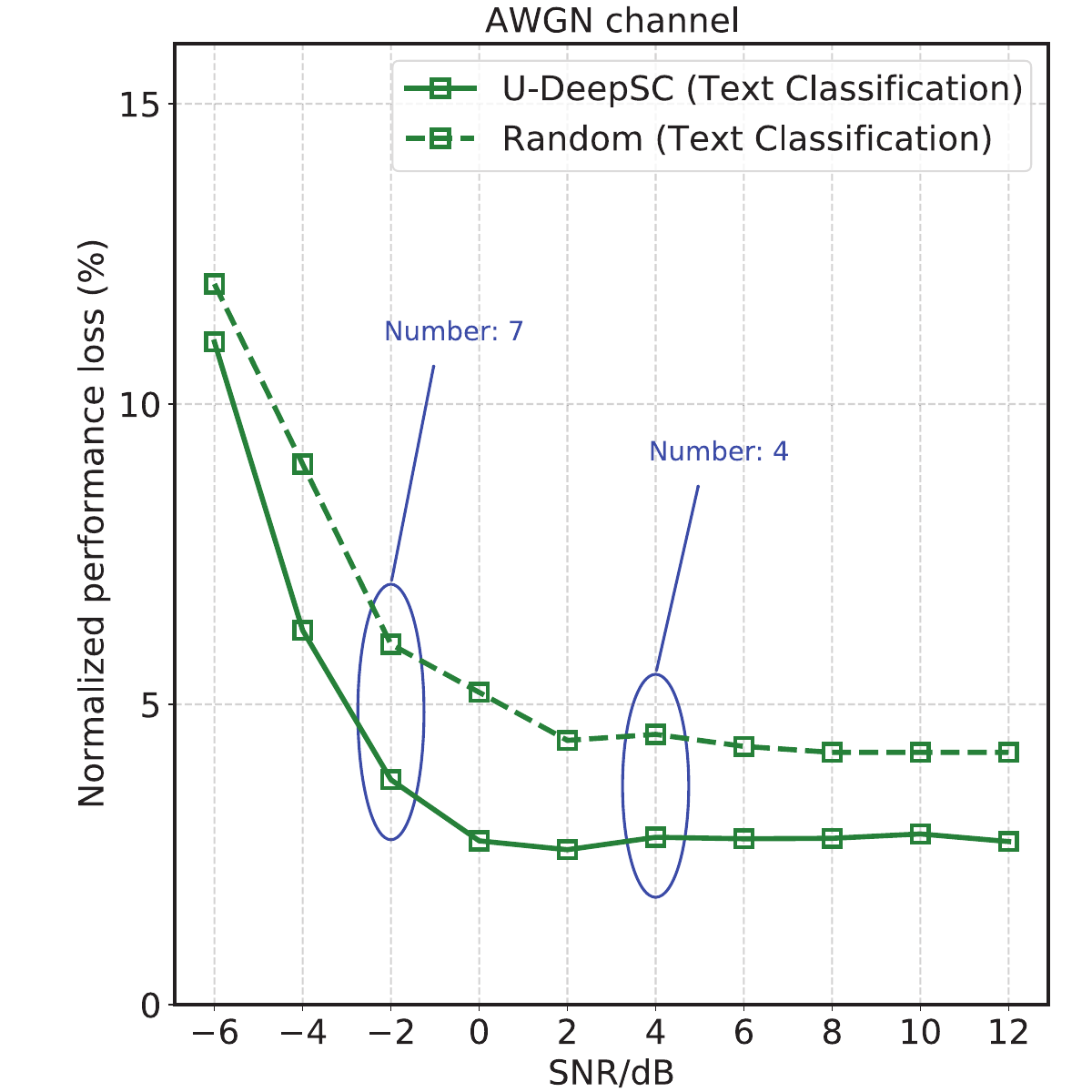}}
			\subfloat[]{\label{overhead2}\includegraphics[width=4.5cm]{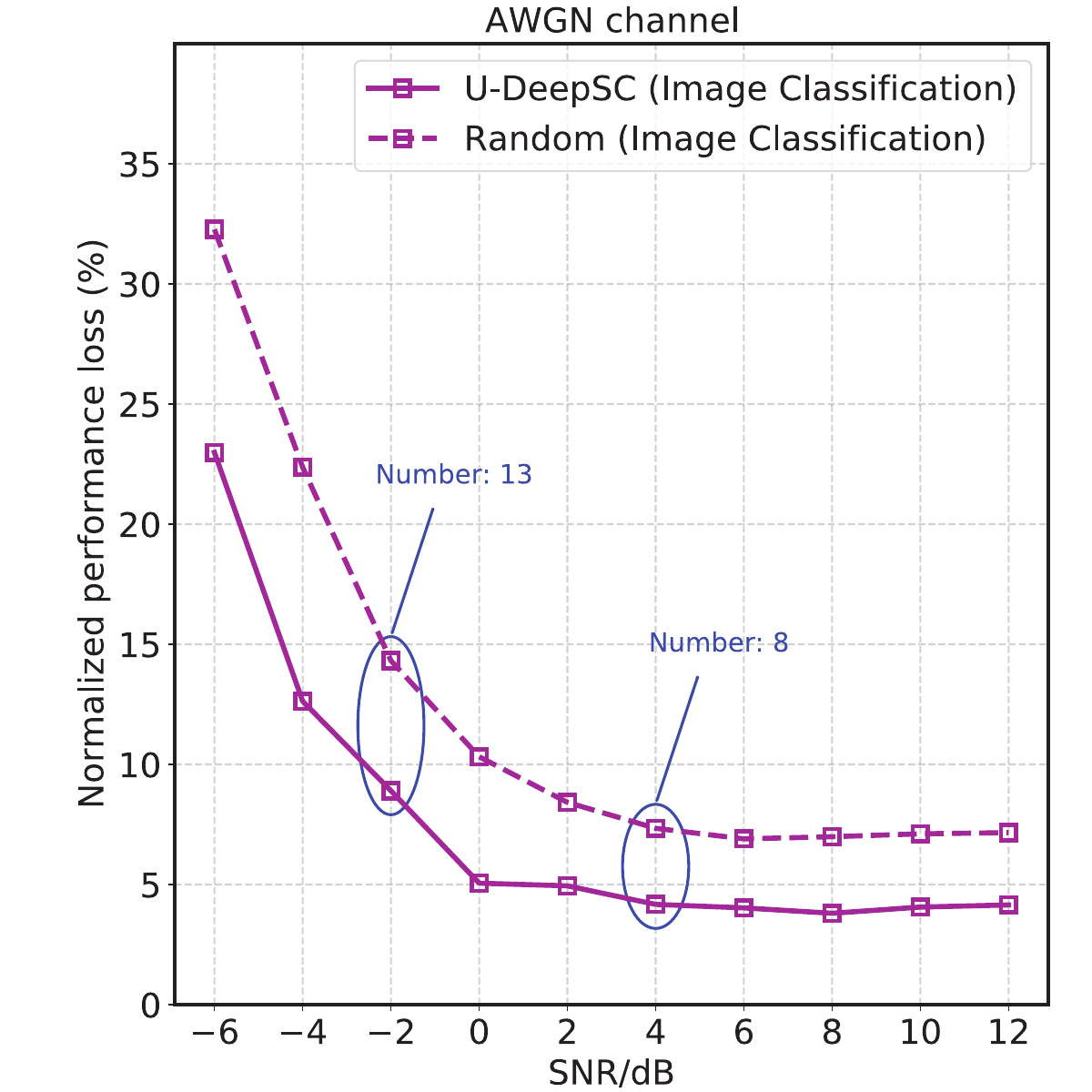}}\quad
			\caption{Performance comparison between FSM and random selection. The "Number " marked in the figure indicates the number of transmitted channel symbols.} 
			\label{Random}
		\end{centering}
	\end{figure}
	
	\subsection{Effectiveness of FSM}
	Fig. \ref{overhead12}(a) and Fig. \ref{overhead12}(b) illustrate the number of transmitted symbols and normalized performance loss versus SNR. From the figure, the normalized performance losses for both U-DeepSC and T-DeepSC decrease with SNR and so does the number of transmitted symbols of U-DeepSC. Compared with T-DeepSC that has a fixed transmission overhead, the proposed method achieves comparable performance with less transmission overhead. The proposed dynamic channel encoder can adaptively adjust the number of transmitted features according to the channel noise levels, thus it can significantly reduce the transmission overhead at the higher SNR regime. Particularly, when the channel conditions are unfavorable, the dynamic channel encoder tends to select more features for transmission, making the received features robust to maintain the performance. It is analogous to adding redundancy for error correction in conventional channel coding techniques. On the contrary, when the channel conditions are good, the dynamic channel encoder tends to transmit fewer features to reduce the communication overhead.

	We conduct feature visualization in Fig. \ref {overhead56}. Note that in the Transformer architecture, each encoded feature corresponds to one small patch of the input image or one word of a sentence. As for the image, we use gray patches to represent the patches that are masked. It is readily seen that the dynamic channel encoder is able to identify the informative features, which are drawn from the important task-specific patches. Moreover, we observe that the patches in the middle of the image have a higher probability to be kept, which is mainly because that in most images, the objects are located in the center. As for the text, the discarded words are marked in red. We can see that the features corresponding to the emotional words, e.g., good and bad, are more likely to be kept in the text classification task. These emotional words usually play an important role in identifying the sentiment of the sentence.

	Next, we verify the effectiveness of our proposed FSM by comparing with the random selection strategy. We employ the model trained at SNR $=12$ dB for this experiment. Specifically, the  feature vectors for proposed FSM are selected based on the probability matrix $\bm{P}$, where only the top $N_d \delta^N_q$ feature vectors with the highest probability will be selected. For comparison, we randomly select $N_d  \delta^N_q$ feature vectors in the inference phase. According to Fig. \ref{Random}, we find that FSM significantly outperforms the random selection strategy. It is mainly because the feature vectors with the higher sampling probability contributes much more  to the final result, i.e., classification results. From the results, we can conclude that FSM makes semantic extraction more efficient and can focus on the content that is truly relevant to the results. Hence, the proposed U-DeepSC is more flexible and can significantly improve the inference speed.

	\subsection{Model Parameters}
	\begin{table}[tb]\footnotesize
		\caption{The number of parameters.}
		\centering
		\begin{tabular}{c|c|c}
				\toprule[0.3mm]
			Task    & T-DeepSC        & U-DeepSC   \\ 
			\midrule[0.18mm]
			\midrule[0.18mm]
			Image  Classification   & 7.65M   & 42.85M  \\ 
			Image Reconstruction       & 8.20M   & 42.85M  \\ 
			Text Classification         & 29.30M    & 42.85M   \\ 
			Text Reconstruction     &33.69M    & 42.85M    \\ 
			Visual Question Answering        & 37.60M   & 42.85M    \\ 
			Video Sentiment Analysis        & 37.23M   & 42.85M   \\ 
			\midrule[0.18mm]
			Stored Parameters         & 153.67M    & 42.85M   \\ 
			\bottomrule[0.3mm]
		\end{tabular}
		\label{table2}
	\end{table}
	As shown in Table \ref{table2}, the total stored number of parameters of T-DeepSC is 153.67M, which is obtained by adding the parameters required for each task. For our proposed U-DeepSC, the number of stored model parameters is only 42.85M for six tasks, which is 28.54\% of the T-DeepSC. The U-DeepSC is able to provide satisfactory performance with much-reduced model parameters. It is of great significance towards a practical semantic communication system for scenarios with limited spectrum and storage resources.

	\section{Conclusion} \label{Conclusion}
	In this paper, we first proposed a general framework for U-DeepSC. Particularly, we considered six popular tasks and jointly trained these tasks with a unified model. To control the transmission overhead, we developed a novel vector-wise FSM to make U-DeepSC adaptive to the tasks, where the number of the transmitted features can be dynamically adjusted for different tasks under different SNR regimes. Then, the unified codebook has been proposed for the feature representation of multiple tasks.  Simulation results showed that our proposed model had satisfactory performance in the low SNR regime and achieved comparable performance to the task-oriented model designed for a specific task with significant reductions in both transmission overhead and model size.

	\bibliographystyle{IEEEtran}
	\bibliography{IEEEabrv,Reference}
	
\end{document}